%% file: main.tex
\documentclass[sigconf]{acmart}
\usepackage{color,soul}

\usepackage{amsmath,amssymb,amsfonts}
\usepackage{bbm}
\usepackage[toc,page]{appendix}
\usepackage{booktabs} 
\usepackage{subcaption} 
\usepackage{graphicx}  	 
\usepackage{lipsum}
\usepackage{float}
\usepackage{enumerate}
\usepackage{enumitem}
\usepackage{url}

\usepackage[ruled,linesnumbered]{algorithm2e} 
\SetAlFnt{\small}

\AtBeginDocument{%
  \providecommand\BibTeX{{%
    \normalfont B\kern-0.5em{\scshape i\kern-0.25em b}\kern-0.8em\TeX}}}

\setcopyright{none}
\renewcommand\footnotetextcopyrightpermission[1]{}
\usepackage{etoolbox}

\acmConference[]{}{}{}

\begin{document}

\title{Genome Reconstruction Attacks Against Genomic Data-Sharing Beacons}

\author{Kerem Ayoz}
\affiliation{%
  \institution{Bilkent University}
  \city{Ankara}
  \country{Turkey}}
\email{keremayoz@hotmail.com}

\author{Erman Ayday}
\authornote{Co-senior  authors.\\ Research reported in this publication was supported by the National Library Of Medicine of the National Institutes of Health under Award Number R01LM013429.}
\affiliation{%
  \institution{Case Western Reserve University}
  \city{Cleveland, OH}
  \country{USA}
  \state{\\and}}
  \affiliation{%
  \institution{Bilkent University}
  \city{Ankara}
  \country{Turkey}}
  \email{exa208@case.edu}

\author{A. Ercument Cicek}
\authornotemark[1]
\affiliation{%
  \institution{Bilkent University}
  \city{Ankara}
  \country{Turkey}
  \state{\\and}}
  \affiliation{%
  \institution{Carnegie Mellon University}
  \city{Pittsburgh, PA}
  \country{USA}}
\email{cicek@cs.bilkent.edu.tr}

\begin{abstract}
Sharing genome data in a privacy-preserving way stands as a major bottleneck in front of the scientific progress promised by the big data era in genomics. A community-driven protocol named \textit{genomic data-sharing beacon protocol} has been widely adopted for sharing genomic data. The system aims to provide a secure, easy to implement, and standardized interface for data sharing by only allowing yes/no queries on the presence of specific alleles in the dataset. However, beacon protocol was recently shown to be vulnerable against membership inference attacks.
In this paper, we show that privacy threats against genomic data sharing beacons are not limited to membership inference. We identify and analyze a novel vulnerability of genomic data-sharing beacons: genome reconstruction. We show that it is possible to successfully reconstruct a substantial part of the genome of a victim when the attacker knows the victim has been added to the beacon in a recent update. In particular, we show how an attacker can use the inherent correlations in the genome and clustering techniques to run such an attack in an efficient and accurate way.
We also show that even if multiple individuals are added to the beacon during the same update, it is possible to identify the victim's genome with high confidence using traits that are easily accessible by the attacker (e.g., eye color or hair type). Moreover, we show how a reconstructed genome using a beacon that is not associated with a sensitive phenotype can be used for membership inference attacks to beacons with sensitive phenotypes (e.g., HIV+). 
The outcome of this work will guide beacon operators on when and how to update the content of the beacon and help them (along with the beacon participants) make informed decisions.
\end{abstract}






\maketitle

\input{sections/intro}
\input{sections/related_work}

\input{sections/preliminary}

\input{sections/model}
\input{sections/proposed_work}

\input{sections/evaluation}

\input{sections/discussion}

\input{sections/conclusion}





\end{document}

%% file: sections/intro.tex
\section{Introduction}\label{section:introduction}
With plummeting sequencing costs, we look forward reaching a capacity of sequencing one billion individuals over the next 15-20 years, resulting in availability of very large genomic datasets~\cite{schatz2015biological,collins2015new,ledford2016astrazeneca}. Although such large datasets are promising a revolution in medicine, it has been shown in numerous studies that it is not straightforward to ensure anonymity of the participants in such datasets~\cite{homer2008resolving,sankararaman2009genomic,jacobs2009new,visscher2009limits,clayton2010inferring}.

Human genome is the utmost personal identifier and sharing genomic data for research while preserving the privacy of the individuals have been challenging many different fields (e.g., medicine, bioinformatics, computer science, law, and ethics) for long, due to possibly dire ethical, monetary, and legal consequences. To address this challenge and create frameworks and standards to enable the responsible, voluntary, and secure sharing of genomic data, the Global Alliance for Genomics and Health (GA4GH) was formed by the community~\cite{ga4gh}. The current genomic data sharing standard of the GA4GH is called the \textit{genomic data-sharing beacons}. Beacons are the gateways that let users (researchers) and data owners exchange information without -in theory- disclosing any personal information.
A user who wants to apply for access to a dataset can learn whether individuals with specific alleles (nucleotides) of interest are present in the beacon through an online interface. That is, a user can submit a query, asking whether a genome exists in the beacon with a certain nucleotide at a certain position, and the beacon answers as "yes" or "no".  If the dataset does not contain the desired genome, genomic data is not shared and distributed unnecessarily. In addition, researchers do not have to go through the paperwork to obtain a dataset which will not be helpful for their research. The GA4GH provides a shared beacon interface~\cite{beacon_network} that as of August 2020 provides access to 80 beacons and acts as a hub where researchers and data owners meet.

Beacons are typically associated with a particular sensitive phenotype (e.g., the SFARI beacon that host individuals with autism). Therefore, presence of an individual in a particular beacon is considered as privacy-sensitive information and the main aim of the beacons is to protect this information. An attacker, using the responses of a beacon and genomic data of a victim, may try to infer the membership of the victim in a particular beacon by running a membership inference attack. Beacon framework sets a barrier against membership inference attacks by allowing only presence/absence queries for variants and not tying any response to any specific individual. In that sense, beacons are considered to have stronger privacy measures compared to other statistical genomic databases. Despite these barriers, several works have proven that beacons are not bulletproof and they are vulnerable to membership inference attacks~\cite{shringarpure2015privacy,Rraisaro2016privacy,vonthenen2018reidentification}.

However, threats against genomic data-sharing beacons are not limited to membership inference attacks. In this paper, for the first time, we identify and analyze the vulnerability of genomic data-sharing beacons for the ``genome reconstruction'' attack. We consider a scenario, in which the attacker knows the membership of a victim to a beacon that may not be associated with a sensitive phenotype. Then, we show how the attacker can accurately infer the genome of the victim by using the beacon responses. Such an attack may result in serious consequences if the attacker uses the reconstructed genome to infer sensitive information (e.g., disease diagnosis) about the victim or to infer the victim's membership to another statistical genomic database of interest (e.g., another beacon that is associated with a sensitive phenotype). In particular, we show how the attacker can use the inherent correlations in the genome to run such an attack in an efficient and accurate way compared to a baseline approach. We also show how clustering techniques can be used to further improve the accuracy of such an attack.

Previous works in the literature assume beacons are static and do not change over time. However, beacons are dynamic datasets (donors join and leave) and this results in an increased risk for the genome reconstruction attack. Thus, for the first time, we consider the beacons as dynamic databases and formulate the genome reconstruction attack accordingly.

In a genome reconstruction attack, the attacker reconstructs all or a subset of the genomes in the beacon. Among the reconstructed genomes, it is not trivial to infer which one belongs to the victim.
Therefore, we also show how the attacker can identify the victim's genome among the set of reconstructed genomes using a set of visible phenotypes (physical characteristics) of the victim, which is public information. Finally, to show one of the consequences of the identified genome reconstruction attack, we show how the attacker can utilize the outcome of this attack to initiate a membership inference attack against the same victim in another beacon, which can be associated with a sensitive phenotype. To do this, we combine the identified genome reconstruction attack with the membership inference attacks against beacons from the literature.

We implement and evaluate the identified vulnerability using real genome data obtained from OpenSNP~\cite{greshake2014opensnp} and HapMap~\cite{international2003international} datasets. We particularly evaluate the success of the attacker to reconstruct a victim's point mutations that include at least one rare nucleotide (i.e., minor allele) since minor alleles (i) reveal sensitive attributes of individuals (e.g., predispositions to privacy-sensitive diseases); and (ii) provide rich information to the attacker for membership inference attacks~\cite{Rraisaro2016privacy,vonthenen2018reidentification}. We show that precision and recall of the reconstruction reach up to 0.9 (each) when 3 individuals are added to the beacon and the victim is one of the newcomers. Even when $10$ new participants are added to the beacon, we show that the attacker has a precision of $0.7$ and a recall of $0.8$.
Furthermore, our results show that when more than one individual is added to the beacon, the attacker can accurately pinpoint the victim's reconstructed genome by matching the victim's phenotypical characteristics to the reconstructed genomes using machine learning algorithms.
We also show via experiments that the outcome of the genome reconstruction attack can be accurately used for the membership inference attack on another beacon and it helps an attacker infer the membership of a victim only with a few queries.

This study clearly shows that privacy risks for genomic data-sharing beacons are much severe than perceived. This is particularly important since the number of beacon participants, and hence the privacy risk of individuals increase rapidly.  
The rest of the paper is organized as follows. In the next section, we summarize the related work in genomic privacy. In Section~\ref{sec:background}, we provide background information about genomics and membership inference attacks against beacons. In Sections~\ref{sec:system_model} and~\ref{sec:threat_model}, we introduce the system and threat models. In Section~\ref{section:proposed_work}, we provide the details of the identified vulnerability. In Section~\ref{section:evaluation}, we evaluate the identified vulnerability using real genomic datasets. In Section~\ref{section:discussion}, we discuss our main findings and potential mitigation techniques. Finally, we conclude the paper in Section~\ref{section:conclusion}.

%% file: sections/related_work.tex
\section{Related Work}\label{section:relatedwork}
Genomic privacy has recently been explored by many studies~\cite{erlich2014routes,survey:genomicera,ayday2013chills}. In the following subsections, we will summarize existing work on privacy in statistical genomic databases, inference attacks, and privacy of genomic data-sharing beacons.

\subsection{Privacy in Statistical Genomic Databases and Inference Attacks on Genomic Privacy}
Several works have shown that anonymization does not effectively protect the privacy of genomic data~\cite{gitschier2009inferential,gymrek2013identifying,Hayden2013,MalinS04,anony:name,lin04,Kale2017AUM}. It has been shown that the identity of a participant of a genomic study can be revealed by using a second sample (e.g., part of the DNA information from the individual) and the results of the clinical study~\cite{related:homer,related:wang,im2012sharing,clayton2010inferring,Zhou_ESORICS_2011}. Differential privacy (DP)~\cite{Dwork:2006:DP:2097282.2097284} concept has been frequently used to mitigate membership inference attacks when releasing summary statistics from genomic databases. Fienberg \emph{et al.} used the DP concept for sharing statistics, such as minor allele frequencies and chi-square values~\cite{differential:gwas}. Yu \emph{et al.} extended this work and presented a scalable algorithm for any arbitrary number of point mutations (single nucleotide polymorphisms - SNPs)~\cite{differential:gwas_yu}. Johnson and Shmatikov proposed using the exponential mechanism for the computation and release of statistics about a genomic database~\cite{differential:gwas_johnson}. Tramer \emph{et al.} also studied the tradeoff between privacy and utility provided by DP~\cite{THHA15}. Compared to statistical databases, genomic data-sharing beacons have stronger privacy measures since they only allow presence/absence (or yes/no) queries for variants.

Humbert \emph{et al.} proposed an inference attack on kin genomic privacy using the family ties between individuals, pairwise correlations between the SNPs, and publicly available statistics about DNA~\cite{genomic:lacks}. Then, Deznabi \emph{et al.} demonstrated that stronger inference techniques can be generated by combining high-order correlations and family ties~\cite{Khodam}. Furthermore, several studies have examined phenotype prediction from genomic data, as a means of tracing identity~\cite{Humbert2015DeanonymizingGD,Lippert201711125,kayser2011improving,zubakov2010estimating,ou2012predicting,allen2010hundreds,manning2012genome,walsh2011irisplex,claes2014modeling,liu2012genome}.
To mitigate such attribute inference attacks, besides DP-based solutions (to release genomic data), cryptographic solutions has been also proposed to perform some operations on genomic data in a privacy-preserving way. Existing cryptographic solutions mainly focus on (i) private pattern-matching and the comparison of genomic sequences~\cite{Pastoriza_CCS_2007,DeCristofaro:2013:SGT:2517840.2517849,related:computation,related:securecomparison,Naveed:2014:CFE:2660267.2660291} and (ii) privacy-preserving personalized medicine~\cite{related:baldi,related:ermanclinic}. In this work, we identify and analyze a different type of attribute inference attack particularly against genomic data-sharing beacons.

\subsection{Privacy in Genomic Data Sharing Beacons}

Researchers showed that presence (membership) of an individual in a genome sharing beacon
can be inferred by repeatedly querying the beacon. Shringarpure and Bustamante introduced a likelihood-ratio test (LRT) that can predict whether an individual is in the beacon by querying the beacon for multiple SNPs of a victim~\cite{shringarpure2015privacy}. Note that inferring the membership of an individual in a beacon that is associated with a sensitive phenotype is equivalent to uncovering the sensitive phenotype about the victim. Then, Raisaro \emph{et al.} showed that if the attacker first queries the SNPs with low minor allele frequency (MAF) values, it needs fewer queries for a successful attack~\cite{Rraisaro2016privacy}. Later, von Thenen \emph{et al.} showed that even if the attacker does not have victim's low-MAF SNPs, it is still possible to infer membership by exploiting the correlations in the genome~\cite{vonthenen2018reidentification}. Furthermore, they showed that beacon responses can also be inferred using such correlations (via a query inference, or QI-attack). In an orthogonal work, Hagestedt \emph{et al.} have hypothesized that while current beacons systems are limited to genomic data, in the near future, the community is going to need a similar system for other biomedical data types. They proposed a beacon system for sharing DNA methylation data (an epigenetic mechanism to regulate transcriptional activity) and then showed that it is possible to successfully launch a membership inference attack against this system. They proposed a DP-based solution in their proposed \textit{MBeacon} system. The approach retains utility by adjusting the noise level for high risk methylation regions that might leak phenotypic information (i.e., regions which are related to disease).

\noindent\textbf{Contribution of this paper.} In this paper, we identify and analyze a genome reconstruction attack against genomic data-sharing beacons by particularly exploiting the information leaked due to beacon updates and the correlations between the point mutations. So far, all works in the literature have focused on membership inference attacks against genomic data-sharing beacons. To the best of our knowledge, this is the first work that identifies, thoroughly analyzes, and shows the consequences of the genome reconstruction attack against the beacons. Furthermore, as opposed to existing work (that only consider a snapshot of the beacon), we show the privacy risk in dynamic beacons, in which new donors may join or existing donors may leave.

%% file: sections/preliminary.tex
\section{Background}\label{sec:background}

In this section, we provide background information on genomics and also on membership inference attack against beacons (that we use in Section~\ref{sec:genome_membership}).

\begin{figure*}[ht]
\centering
\includegraphics[width=0.80\textwidth]{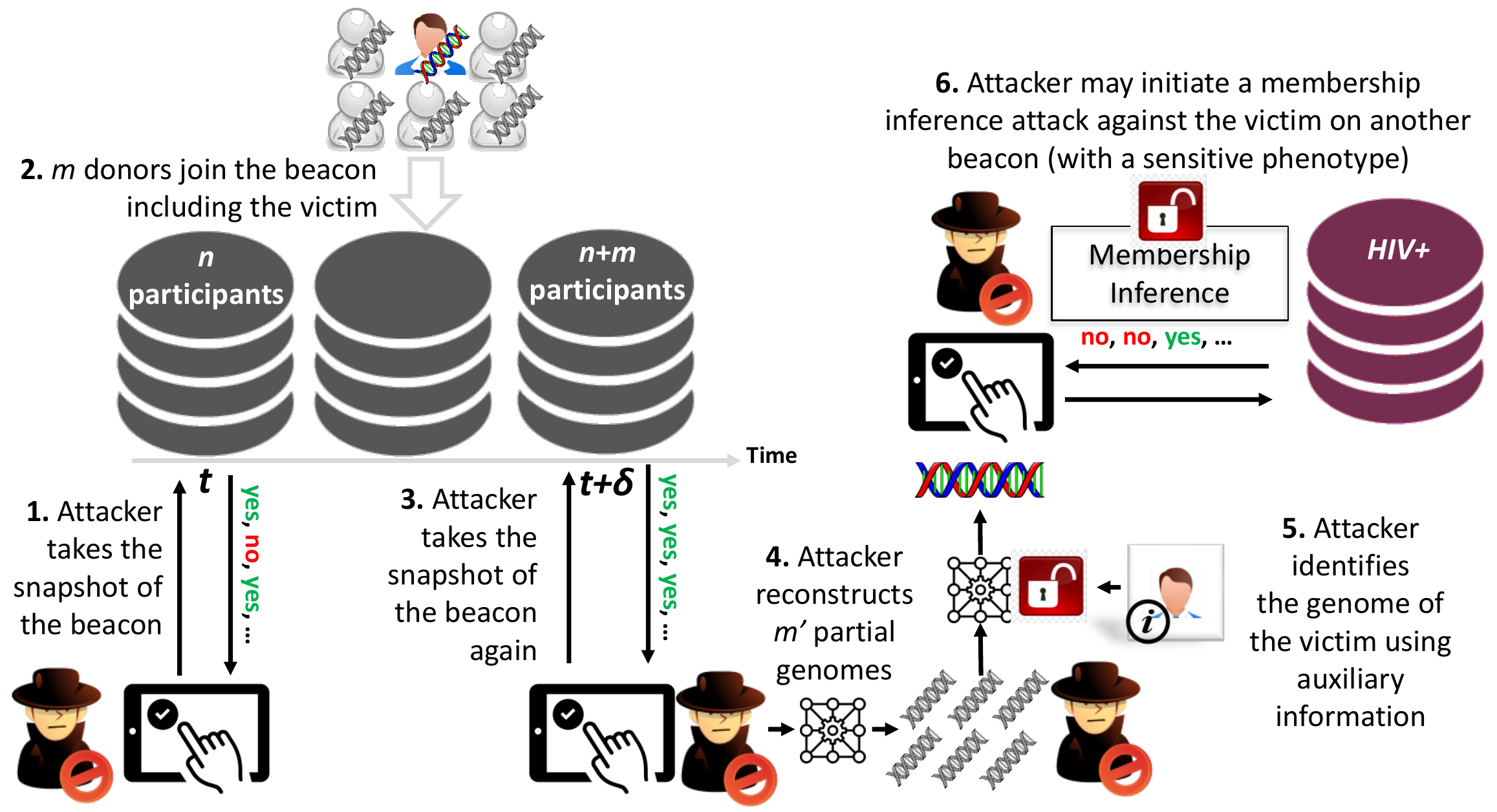}
\footnotesize
\caption{Proposed system model.}
\label{fig:system_model}
\end{figure*}
\subsection{Genomics Background }\label{sec:geneticbackground}

Approximately $99.9\%$ of the all individuals' DNA are identical and the remaining $0.1\%$ is responsible for our differences.
Single nucleotide polymorphism (SNP) is the most common source of variation in the human genome. SNP is a point mutation (e.g., substitution of a single nucleotide in the genome - A,T,C, or G) and there are around 50 million known SNPs in the human genome~\cite{site:snppop}. The alternative nucleotides for each locus (SNP position) are called alleles and each allele of a SNP can be either the major or the minor allele for that SNP. The major allele is the most frequently observed nucleotide for a SNP position and the minor allele is the rare nucleotide (i.e., the second most common). The frequency (or probability) of observing the minor allele at a SNP position is called the minor allele frequency (MAF) of that SNP. Human genome has two copies for each locus (one per chromosome) and a SNP can be represented in terms of the number of its minor alleles (i.e., $0$ for homozygous major, $1$ for heterozygous, or $2$ for homozygous minor).

Recent discoveries show that particular SNPs in human population are inherently correlated and this correlation model may change for different populations.
Linkage disequilibrium (LD) is the non-random association of alleles at two or more loci. If two SNPs are in LD, they are correlated and co-occur more frequently than expected. Some SNPs are pathogenic and cause genetic diseases~\cite{site:snpdiseases} and hence, they may carry sensitive information regarding individuals' health conditions. As discussed in Section~\ref{section:relatedwork}, most existing works in genomic privacy literature focus on the protection of the SNPs to prevent the risk of genetic discrimination.

\subsection{Membership Inference Attack Against Genomic Data-Sharing Beacons}\label{sec:membership_inference}
In~\cite{Rraisaro2016privacy}, Raisaro \emph{et al.} introduced the Optimal attack, in which the attacker constructs a set of candidate SNPs $S$ to be queried and submits queries starting from the lowest MAF $SNP_i$.
Let the null hypothesis ($H_0$) refer to the case in which the queried genome is not in the beacon and alternative hypothesis ($H_1$) be the case in which the queried genome is a member of the beacon.
In~\cite{Rraisaro2016privacy}, the log-likelihood ($L$) under the null and alternate hypothesis are shown as follows:
\begin{equation} \label{eq:QIHA}
\begin{split}
L_{H_{0}}(R)  =  \sum_{i=1}^{n} x_{i}\text{log}(1-D^i_{N}) +(1-x_{i})\text{log}(D^i_{N})
\end{split}
\end{equation}
\begin{equation} \label{eq:QIHB}
\begin{split}
L_{H_{1}}(R)  =  \sum_{i=1}^{n} x_{i}\text{log}(1-\delta D^i_{N-1}) +(1-x_{i})\text{log}(\delta D^i_{N-1}),
 \end{split}
\end{equation}
where $R$ is the response set, $x_i$ is the answer of the beacon to the query at position $i$ (1 for ``yes'', 0 for ``no''), and $\delta$ represents a small probability where the attacker's copy of the victim's genome does not match the beacon's copy for a locus (e.g., due to difference in variant calling pipeline). $n$ is the number of posed queries. $D_{N}^i$ is the probability that none of the $N$ individuals in the beacon has the queried allele at position $i$ and $D_{N-1}^i$ represents the probability of no individual except for the queried person having the queried allele at position $i$. The computations of $D_{N-1}^i$ and $D_{N}^i$ depend on the queried position $i$ and they change at each query as follows: $D_{N-1}^i=(1-f_i)^{2N-2}$ and $D_{N}^i=(1-f_i)^{2N}$, where $f_i$ represents the MAF of the SNP at position $i$. The likelihood-ratio test (LRT) statistic, $\Lambda$, is then determined as
\begin{equation}\label{eq:QILambda}
\begin{split}
\Lambda = \sum_{i=1}^{n} \text{log}\Biggl(\frac{D^i_N}{\delta D^i_{N-1}}\Biggl)+\text{log}\Biggl(\frac{\delta D^i_{N-1}(1-D^i_N)}{D^i_N(1-\delta D^i_{N-1})}\Biggl) x_i.
\end{split}
\end{equation}

In Section~\ref{sec:genome_membership}, we use the Optimal attack when we show how the proposed genome reconstruction attack can be combined with the membership inference attack.

%% file: sections/model.tex
\section{System Model}\label{sec:system_model}

As shown in Figure~\ref{fig:system_model}, we consider a system between the beacon participants (e.g., donors), the beacon, and the beacon users. The donor shares their genome with the beacon. It is possible that the donor may share their genome with more than one beacons that may or may not be associated with sensitive traits. Genome donor is not active during the protocol after they share their data with the beacon. Also, beacon never publicly shares its dataset. Some beacons may only share metadata about their donors such as their gender, age, or ethnicity. In general, we consider the beacon as a dynamic dataset in which new donors may join and existing donors may leave over time. Beacon users issue queries to the beacon. A beacon user is a potential attacker as shown in Figure~\ref{fig:system_model}. As discussed, the beacon user can only ask the presence of a genome with a particular allele (nucleotide) at a particular position of a given chromosome and the beacon only responds as ``yes'' or ``no''. In this work, we assume beacon honestly reports the result of each query to the user (e.g., without introducing intentional noise to the query results) and we do not consider a query limit for the users as it is usually trivial to overcome such limits (e.g., by registering several times with different accounts).

\section{Threat Model}\label{sec:threat_model}

Depending on the attacker's objective, two attacks that can be launched against genomic data-sharing beacons are: (i) membership inference attack and (ii) genome reconstruction attack. Here, for the first time, we identify and study the latter. We assume that the attacker, with the knowledge about the membership of an individual to a beacon, tries to reconstruct a victim's genome by issuing queries to the corresponding beacon. This is a realistic assumption, especially for beacons that are not associated with a sensitive trait (e.g., Kaviar~\cite{glusman2011kaviar}).
For such beacons, membership of an individual may not be privacy-sensitive information. However, using this information, the attacker may infer the genome of the victim and use this for other attacks against the victim.

This vulnerability exists both for static and dynamic beacons. In static beacons, knowing that the victim is a member of the beacon, only the ``no'' responses would provide certain information about the victim's genome to the attacker. ``Yes'' responses may be due to any other participant of the beacon and as the size of the beacon increases, ``yes'' responses do not provide much information to the attacker. However, in dynamic beacons, when the beacon is updated, using the change in the responses of the beacon, the attacker can learn more about the genomes of new participants. Thus, in this paper, we analyze this vulnerability for dynamic beacons and we assume that the victim is added between times $t$ and $t+\delta$ along with other $(m-1)$ newly added donors to the beacon.

We assume that, along with the fact that the victim is among the newly joined participants to the beacon, the attacker also knows (i) the number of other newly joined individuals that are added to the beacon along with the victim;
(ii) a snapshot of the beacon before the victim is added (at time $t$). That is, responses to all queries before the victim joins to the beacon; (iii) a set of victim's visible characteristics (phenotype); and (iv) publicly available information about genomics, such as minor allele frequencies (MAF values) of SNPs and correlation between the SNPs in the population of interest. Finally, we assume that the attacker is a regular beacon user and it does not collude with the beacon. 

%% file: sections/proposed_work.tex
\section{Genome Reconstruction Attack on Genomic Data-Sharing Beacons}\label{section:proposed_work}

As discussed, we define the genome reconstruction attack as inferring genomic data of a genome donor (i.e., victim) given their membership information to the beacon. To show the effect of genome reconstruction attack more clearly, we consider dynamic beacons and we assume the victim is among the newly joined donors to the beacon. For clarity of the discussion, we present the identified attack only considering newly joined donors. Considering the donors that leave the beacon is symmetrical and trivial. We discuss this case in Section~\ref{sec:leaving_donors}.

In genome reconstruction attack, due to the nature of beacon responses, the attacker can infer if a victim has at least one minor allele at every SNP position. This is because the response of the beacon only tells if there is an individual in the beacon with at least one minor allele at a given SNP position. Thus, for each SNP $j$ of victim $v$ ($S_j^v$), the goal of the attacker is to infer $Pr(S_j^v=0)$ and $Pr(S_j^v\neq0)$ (i.e., $Pr(S_j^v=1)$ or $Pr(S_j^v=2)$).
For simplicity, we define the event
$\hat{S}_j^v=\mathbbm{1}_{S_j^v=1 \lor S_j^v=2}$.
Thus, $\hat{S}_j^v=0$ if $S_j^v=0$, and $\hat{S}_j^v=1$, otherwise.
Note that inferring this information for a victim results in a serious privacy concern. As we will discuss and show later, using this information, an attacker can associate the genotype of the victim to related phenotypes (e.g., diseases) and initiate a membership inference attack for the victim by targeting another beacon that is associated with a sensitive phenotype (e.g., cancer or HIV+).

We consider a scenario in which the attacker has no information about the victim's genome, but it knows that the victim is added to the beacon between times $t$ and $t+\delta$. Let $n$ and $(n+m)$ represent the number of individuals in the beacon at times $t$ and $t+\delta$, respectively. We also assume that the attacker knows $m$, which can easily be obtained by monitoring the changes in beacon size (or from the metadata of the beacon).
By possessing this information, the attacker can probabilistically infer the genome of the victim by utilizing the changes in beacon's responses (at times $t$ and $t+\delta$) as follows: (i) if the previous response (at time $t$) was ``no'' and the current response (at time $t+\delta$) is ``yes'', the probability that the victim having a minor allele at the corresponding query position increases depending on how many new individuals are added to the beacon in this time interval; (ii) if the previous response was ``yes'' and the current response is also ``yes'', attacker cannot infer much about the victim's genome, especially if the total size of the beacon is large; and (iii)
if both the previous and the current responses are ``no'', the attacker understands that the victim does not have a minor allele at the corresponding query position.

Here, the most important (or the most sensitive) information for the attacker can be considered as the ``no'' responses at time $t$ that turn to ``yes'' at time $t+\delta$. Because, such responses let the attacker infer the positions that the victim has at least one minor allele with a high probability (depending on how many new individuals are added to the beacon in this time interval). Since minor alleles of individuals are typically the indicators for privacy-sensitive information about them, in this work, we focus on the success of the attacker based on its success in inferring the minor alleles of a victim using the beacon responses that turn to ``yes''. Exhaustively generating all potential solutions of this problem would result in a total of $2^{\beta*m}$ genomes, where $\beta$ is the total number of responses that turn to ``yes'' at time $t+\delta$ (which can be on the order of tens of thousands), and hence it is intractable. In the following, we first describe a baseline method that provides a tractable solution to this problem. Next, we present a greedy approach to run such an attack more accurately, and then we will detail a more sophisticated, clustering-based approach for the genome reconstruction attack.

\subsection{Baseline Approach for Genome Reconstruction}\label{sec:baseline}

Here, we describe a baseline approach, in which the attacker, using the responses of the beacon, reconstructs the genomes (of the newly joined donors) by assigning them to $m'$ bins according to MAF values of the SNPs ($m'$ can be different than $m$ and the selection of $m'$ effects the precision and recall of the attacker). Genome reconstruction attack using the baseline algorithm for a particular victim $v$ at time $t+\delta$ can be described as follows. The input of the attacker is (i) complete snapshot of the beacon with $n$ donors at time $t$ (ii) the fact that $m$ new donors are added to the beacon between times $t$ and $t+\delta$; (iii) the fact that the victim is among the newly added donors; and (iv) publicly available MAF values of the SNPs.

First, the attacker identifies the set of SNPs for which the response of the beacon was ``no'' at time $t$ and it becomes ``yes'' at time $t+\delta$. Thus, the attacker constructs a set $R_{N\rightarrow{Y}}$, consisting of these SNPs. Then, the attacker creates $m'$ empty bins representing SNP sets of newcomer donors. For each SNP $j$ in set $R_{N\rightarrow{Y}}$, the attacker retrieves its MAF value, $MAF_j$. Next, the attacker assigns the value of SNP $j$ for each individual $i$ (in each bin) consistent with the SNP's MAF value as follows: (i) $\hat{S}_j^i=0$ with probability $(1-MAF_j)^2$ and (ii) $\hat{S}_j^v=1$ with probability $MAF_j^2 + 2MAF_j(1-MAF_j)$.
Since the beacon's response for SNPs in $R_{N\rightarrow{Y}}$ has flipped from ``no'' to ``yes'', for all SNPs in $R_{N\rightarrow{Y}}$, there should be at least one bin (among $m'$ bins) with at least one mutation (i.e., homozygous minor or heterozygous SNP). Thus, once the values of the SNPs in $R_{N\rightarrow{Y}}$ for all $m'$ bins are determined, the attacker checks if there is any SNP in set $R_{N\rightarrow{Y}}$ that is not assigned to any bin. If there is such a SNP, the attacker randomly picks a bin and assigns the value of the corresponding SNP as $\hat{S}_j^i=1$ for the corresponding bin. The details of this baseline approach are also shown in Algorithm~\ref{alg:alg0}.
\input{sections/baseline_alg.tex}

\subsection{Greedy Algorithm for Genome Reconstruction}\label{sec:greedy}

The above-mentioned baseline algorithm assumes every SNP is independent and the correlations among them are disregarded. However, SNPs are inherently correlated and considering such correlations in the genome reconstruction attack may result in significantly more accurate results.
In the greedy algorithm discussed here, the attacker forms the bins considering the correlations between the SNPs in set $R_{N\rightarrow{Y}}$. Using an iterative approach, the attacker assigns each SNP (minor allele) to an individual such that the probability of assignment is proportional to the average correlation of the new SNP with the already assigned SNPs of the individual (i.e., bin $i$). If no assignment is made this way, a random individual is selected to make sure there is at least one person with the corresponding new SNP.

Genome reconstruction attack using the greedy algorithm for a particular victim $v$ at time $t+\delta$ can be described as follows. The input of the attacker is (i) responses of the beacon to all possible queries at time $t$; (ii) the fact that $m$ new donors are added to the beacon between times $t$ and $t+\delta$ and the victim is among the newly added donors; (iii) publicly available MAF values of the SNPs; and (iv) a correlation model between the SNPs that is consistent with the population structure of the beacon (that can be computed using publicly available genomic datasets).

For the correlation model, we assume the attacker uses a Markov chain model, as described in~\cite{samani2015quantifying}. The attacker calculates the likelihood of the victim $v$ having at least one minor allele at a SNP position $j$ as
\begin{equation} \label{eq:MMC}
P_k(\hat{S}_j^v) = P(\hat{S}_j^v|\hat{S}_{j-1}^v, \hat{S}_{j-2}^v, ..., \hat{S}_{j-k}^v),
\end{equation}
where $k$ is the order of the Markov chain. In order to build a Markov chain model for the genome, we use public sources such as HapMap~\cite{gibbs2003international}.
Consistent with the previous work in~\cite{samani2015quantifying}, we define the $k^{th}$-order model as follows: (i)  $P_k(\hat{S}_j)=0$ if $F(\hat{S}_{j-k,j-1}) = 0$ and (ii) $P_k(\hat{S}_j)=\frac{F(\hat{S}_{j-k,j})}{F(\hat{S}_{j-k,j-1})}$ if $F(\hat{S}_{j-k,j-1}) > 0$,
where $F(\hat{S}_{j,i})$ is the frequency of occurrence of the sequence that contains $\hat{S}_i$ to $\hat{S}_j$. The SNPs are ordered according to their physical positions on the genome.
In this work,
we use $k=1$
and we do not limit the correlations only for the neighboring SNPs which is different from~\cite{samani2015quantifying}. Instead, we create our correlation model by considering the pairwise correlations between all the SNPs in the beacon. Here, we use Sokal-Michener distance to measure correlations between SNPs.

In the greedy approach, first, the attacker constructs set $R_{N\rightarrow{Y}}$. Then, it creates $m'$ empty bins ($m'$ does not have to be equal to $m$) representing the number of rare SNPs in $R_{N\rightarrow{Y}}$.
We assume that the SNPs with an MAF value below a threshold $\tau$ are categorized as rare SNPs.
Observing rare SNPs do not have correlations among each other, assigning the rare SNPs in $R_{N\rightarrow{Y}}$ to different bins as seeds is assumed to result in an accurate initial separation of individuals. Next, for each remaining SNP $j$ in $R_{N\rightarrow{Y}}$, the attacker computes the average correlation between that and all the previously assigned SNPs in bin $i$ using the aforementioned correlation model. This is done for each bin $i$. Let $\hat{S}_j^i$ be a binary random variable for SNP $j$ and bin $i$. The attacker assigns $\hat{S}_j^c=1$ for bin $c$ which has the highest average correlation value and $\hat{S}_j^i = 0, \forall i \in [1,m']$ and $i \neq c$. Eventually, the attacker constructs $m'$ potential genomes (in $m'$ bins) belonging to $m$ newcomer donors.

\subsection{Clustering-Based Algorithm for Genome Reconstruction}\label{sec:clustering}

Greedy algorithm (in Section~\ref{sec:greedy}) reconstructs genomes by following a particular order (determined based on the MAFs of the SNPs). Different orders may provide different (and possibly more accurate) solutions. Thus, to consider all query responses together in a collective way, we propose clustering-based approaches for the genome reconstruction attack that cluster the identified minor alleles for the newly joined donors to the beacon. The proposed clustering techniques essentially use the correlations between the SNPs (that are computed using the aforementioned correlation model) to distribute SNPs into different bins. We use two types of clustering techniques: (i) one that creates non-overlapping bins (hard clustering) and (ii) one that may assign a SNP into multiple bins (soft or fuzzy clustering).

For (i), we employ spectral clustering, in which a standard clustering method (such as k-means clustering) is applied on certain eigenvectors of the Laplacian matrix of a graph~\cite{ng2002spectral}.
Spectral clustering is our method of choice as it has been shown to provide favorable results in many high dimensional feature spaces like ours~\cite{rodriguez2019clustering}.
And, for (ii) we employ the fuzzy c-means clustering (FCM) algorithm~\cite{bezdek1984fcm}, which is a common choice for these types of tasks. The algorithm is similar to k-means clustering, but it also allows probabilistic assignments of samples to multiple clusters.
The description of both clustering methods are similar except for the clustering steps. Thus, in the following, we describe both methods together.

The input of both clustering-based algorithms is the same as the input of the greedy algorithm.
First, the attacker identifies the set of SNP positions for which the response of the beacon was ``no'' at time $t$ and it becomes ``yes'' at time $t+\delta$ and constructs set $R_{N\rightarrow{Y}}$.
Then, the attacker builds a graph of SNPs using the correlation model, in which the vertices are the SNPs in $R_{N\rightarrow{Y}}$ and undirected edges are weighted by the correlation values between these SNPs.
This graph represents a pairwise similarity model for the SNPs and is used for a quantitative assessment of the correlation of each SNP pair in $R_{N\rightarrow{Y}}$.

Next, the attacker applies either the spectral or fuzzy clustering algorithms on the constructed graph. The outcome of spectral clustering is a set of disjoint clusters.
Fuzzy clustering results in groups of SNPs that maximizes the similarity in a group while allowing a SNP to be shared by multiple individuals.
Thus, in fuzzy clustering, each SNP $i$ is assigned to clusters for which the algorithm returns a relatively high probability of association.
After clustering, the attacker obtains $m'$ different clusters which corresponds to $m'$ reconstructed genomes.
The details of this algorithm are also shown in Algorithm~\ref{alg:alg1}.
\input{sections/clustering_alg.tex}

\subsection{Identifying the Victim Using Genotype-Phenotype Associations}\label{sec:phenotype}

In previous sections, for genome reconstruction, we assumed that the attacker can correctly identify the victim's genome among several reconstructed bins. Assuming the attacker has information about some phenotypic characteristics of the victim and relying upon the fact that SNPs are intrinsically linked to phenotypic traits (such as eye color, hair color, etc.), we also study and show how accurately the attacker can identify the victim's genome among other candidates. This provides a complete methodology for the genome reconstruction attack against beacons in real-life.

Assume victim $v$ is among the $m$ new additions to the beacon (it is trivial to extend the methodology if there are more than one victims). The attacker is assumed to have access to two distinct sets: (i) a set $\mathcal{S} = \{\vec{S}_1, \vec{S}_2, \ldots, \vec{S}_{m'}\}$ of $m'$ reconstructed genotypes as a result of the genome reconstruction attack, where $\vec{S}_i = (\hat{S}_1^i,\dots,\hat{S}_k^i)$ is a vector containing the SNP values of genotype $i$ (or bin $i$); and (ii) a set $\mathcal{P}_v = (p_{1}^v,\dots,p_{t}^v)$ containing the values of $t$ phenotypic traits of victim $v$. Such phenotype information can be obtained from publicly available resources or using the physical traits of the victim. For instance, the attacker can obtain such information from victim's social media accounts. The goal of the attacker is to correctly match the victim's phenotype to the correct reconstructed genome (that is the most similar to the victim's) among all candidate reconstructed genome sequences.
In the test phase, the attacker has $m$ newly added donors and $m'$ reconstructed genomes. Attacker's task is to match each donor with the best matching reconstructed genome. Thus, for each newly added donor, the attacker calculates the likelihood scores of matching with all $m'$ reconstructed genomes.
In~\cite{humbert2015anonymizing}, Humbert \emph{et al.} focused on the deanoymization risk and modelled genotype-phenotype association as an assignment problem. They showed this risk by using the Hungarian algorithm~\cite{hungarian}.
Different from~\cite{humbert2015anonymizing}, here, we rely on machine learning for maximizing the matching likelihood and genotype-phenotype associations. We observe that such a formulation provides more accurate results. Also, rather than using SNP values (0, 1 or 2), due to the nature of the proposed attack, we represent the state of each SNP $j$ of individual $i$ as $\hat{S}_j^i$, which can be either $0$ or $1$, as discussed before.

For phenotype inference, we train a separate model for each of the considered phenotypes, where SNPs with flipped responses (from ``no'' to ``yes'') are used as features. Since phenotype datasets are highly imbalanced, we apply Synthetic Minority Oversampling Technique (SMOTE)~\cite{smote} for each of these datasets to resolve this problem. Then, we train an RF model for each phenotype. We use repeated stratified 5-fold cross validation to tune the hyperparameters.
After training the phenotype models, we form the ensemble classifier using the ones that have better validation F1-macro score than random guess. We discard the other models.

Ensemble classifier calculates the matching likelihood of given genome and set of phenotypic traits. Softmax output of each phenotype model corresponding to a given phenotypic trait of the victim (i.e., probability that a reconstructed genome having blue eye) are summed to calculate the matching likelihood. For single victim, this calculation is done for each reconstructed genome and the victim is matched with the reconstructed genome with the highest matching likelihood score. Note that this matching does not need to be one-to-one; a single reconstructed genome might match with different set of phenotypic traits. We discuss the performance of identification of victim's reconstructed genome under different settings in Section~\ref{sec:eval_phenotype}.

\subsection{Using Genome Reconstruction in Membership Inference Attack}\label{sec:genome_membership}

To show one consequence of the proposed genome reconstruction attack, we also model and analyze how the proposed attack can be utilized for membership inference attack (introduced in Section~\ref{sec:membership_inference}). We consider a scenario in which the attacker knows the membership of an individual to a beacon with which no sensitive associated phenotype (e.g., phenotype neutral). The attacker first utilizes the responses of this beacon to infer specific parts of a victim's genome (i.e., SNPs). Then, it uses these inferred SNPs to infer the membership of the victim to a beacon with a sensitive phenotype.
This attack is important and realistic, because knowing the membership of an individual to a phenotype neutral beacon (e.g., Kaviar Beacon) may not seem to pose a privacy issue. However, using the proposed genome reconstruction attack and the membership information of the victim to the beacon with non-sensitive phenotype, the attacker can first infer the SNPs of the victim and then, infer the membership of the victim to another beacon which is associated a sensitive phenotype (e.g., SFARI beacon which is associated with autism phenotype).

To show this, first, we run the proposed genome reconstruction attack that is explained in Section~\ref{sec:clustering} and infer the SNPs of the victim with at least one minor allele on a beacon $B_1$. Using these inferred SNPs, we then run the membership inference attack to infer the membership of the victim in another beacon $B_2$.
For membership inference attack, we use the Optimal attack in~\cite{Rraisaro2016privacy} (described in Section~\ref{sec:membership_inference}), which is shown to be an effective attack for membership inference (for our scenario, Optimal attack in~\cite{Rraisaro2016privacy} and the QI-attack in~\cite{vonthenen2018reidentification} perform similarly, so we choose to use the Optimal attack due to its simplicity). However, in contrast to the original Optimal attack, in the null and alternate hypothesis equations in (\ref{eq:QIHA}) and (\ref{eq:QIHB}), there is an additional error due to the inference error of the genome reconstruction attack. This is because the attacker queries the alleles of the victim that it infers as a result of the genome reconstruction attack and there is a degree of uncertainty.
Thus, we first experimentally compute the error rate of the genome reconstruction attack for a particular scenario (e.g., for particular $m$ and $n$ values). We then include this additional error on the $\delta$ parameter in (\ref{eq:QIHB}), which represents the probability that the attacker's copy of the victim's genome does not match the beacon's copy for a SNP.
Furthermore, as opposed to original Optimal attack, here the attacker may not have access to the SNPs of the victim with the lowest MAF values; instead the attacker only knows the SNPs that are inferred as a result of the genome reconstruction attack.

We evaluate the success of this attack in terms of the power of the attacker in Section~\ref{sec:eval_membership}. Similar to~\citeauthor{Rraisaro2016privacy} and~\citeauthor{vonthenen2018reidentification}, we plot the power curve of the membership inference attack at $5\%$ false positive rate. We empirically build the null hypothesis ($H_0$ in Section~\ref{sec:membership_inference}). For every query, we determine the distribution of $\Lambda$ under the null hypothesis using $20$ individuals that are not in $B_2$. In this work, in order to model the uncertainty of correctly matching the victim (using phenotype inference as in Section~\ref{sec:phenotype}), we first experimentally compute the error rate of the overall process. For instance, if the accuracy of correctly matching the phenotype of the victim to their reconstructed genome is $p\%$, then $p\%$ of the $20$ individuals are selected from correctly identified reconstructions and remaining individuals are selected from other new people added to the beacon along with the victim (incorrect identifications).

When $\Lambda$ is less than a threshold $t_\alpha$, the null hypothesis is rejected and we find $t_{\alpha}$ from the null hypothesis with $\alpha=0.05$ (corresponding to $5\%$ false positive rate). Then, we computed the power as proportion of the individuals in the alternate hypothesis (including $20$ different individuals in $B_2$) having a $\Lambda$ value that is less than $t_{\alpha}$. As before, $p\%$ of the $20$ individuals are selected from correctly identified reconstructions and remaining people are selected from other new people added to the beacon along with the victim.

%% file: sections/baseline_alg.tex
\begin{algorithm} [hbt!]

\SetAlgoLined
\DontPrintSemicolon

\KwIn{$b$: beacon; $m$: Number of added people to $b$; Population $P$ that represent the composition in $b$
}
\KwOut{$m'$ reconstructed genomes}
\;
\tcp{Step 1: Query Beacon}
snapshot1 $ \leftarrow  queryBeacon(b, t) $ \;
\tcp{Including victim, $m$ donors join Beacon between time $t$ and $t+\delta$}
snapshot2 $ \leftarrow queryBeacon(b, t+ \delta)$ 
\;
\;
\tcp{Step 2: Obtain No-Yes SNPs}
\SetKw{KwBy}{by}
NoYesResponses $ \leftarrow [] $  \;
\For{$i\gets0$ \KwTo $snapshot1.length$}{
   \If{snapshot1[i] = "No" and snapshot2[i] = "Yes"}{
       NoYesResponses.append(i)\;
   }
 }
\;

\tcp{Step 3: Reconstruct genomes}
S  $ \leftarrow$ []\;
\For{$i \gets 0$ \KwTo NoYesResponses.length}{
    $s \leftarrow NoYesResponses[i]$\;
    $assigned \leftarrow False$\;
    \For{$j \gets 0$ \KwTo $m'$}{
        $S[j][s] \leftarrow  getMajorAllele(P,s)$
        $randnum \leftarrow Random(0,1)$\;
        \If{randnum < getMAF(P,NoYesResponses[i])}{
            $S[j][s] \leftarrow  getMinorAllele(P,s)$\;
            $assigned \leftarrow True$
        }
    }
}
\;
\tcp{Step 4:If a SNP is unassigned, randomly assign it to a reconstruction}
 \If{!assigned}{
    $randnum \leftarrow Random(0,m')$\;
    $S[randnum][s] \leftarrow  getMinorAllele(P,s)$
  }
\;
	
\Return S \;
\caption{Baseline Algorithm for Genome Reconstruction Attack}
\label{alg:alg0}
\end{algorithm}

%% file: sections/clustering_alg.tex
\begin{algorithm} 

\SetAlgoLined
\DontPrintSemicolon

\KwIn{$b$: beacon; $m$: Number of added people to $b$; Population $P$ that represent the composition in $b$
}
\KwOut{$m'$ reconstructed genomes}
\;
\tcp{Step 1: Query Beacon}
snapshot1 $ \leftarrow  queryBeacon(b, t) $ \;
\tcp{Including victim, $m$ donors join Beacon between time $t$ and $t+\delta$}
snapshot2 $ \leftarrow queryBeacon(b, t+ \delta)$
\;
\;
\tcp{Step 2: Obtain No-Yes SNPs}
\SetKw{KwBy}{by}
NoYesResponses $ \leftarrow [] $  \;
\For{$i\gets0$ \KwTo $snapshot1.length$}{
   \If{snapshot1[i] == "No" and snapshot2[i] == "Yes"}{
       NoYesResponses.append(i)\;
   }
 }
\;

\tcp{Step 3: Cluster No-Yes SNPs}
$G \leftarrow Graph()$\; 
 \For{$i\gets0$ \KwTo $NoYesResponses.length - 1$}{
    \For{$j \gets i+1$ \KwTo NoYesResponses.length}{
      $c \gets corr(P,NoYesResponses[i],NoYesResponses[j])$\\
      $G$.addEdge(NoYesResponses[i],NoYesResponses[j],$c$)
    }
  }
$clusters \leftarrow graphClustering($G$, m')$ \;
\;
\tcp{Step 4: Reconstruct genomes}
S  $ \leftarrow$ []\;
\For{$i \gets 0$ \KwTo $m'$}{
	$S[i]  \leftarrow getReferenceGenome(P) $\;
	\ForEach{s in $clusters[i]$}{
		$S[i][s] \leftarrow  getMinorAllele(P,s)$ \;
 	}
}
\;

\Return S \;
 \caption{Clustering-Based Algorithm for Genome Reconstruction Attack}
\label{alg:alg1}
\end{algorithm}

%% file: sections/evaluation.tex
\section{Evaluation}\label{section:evaluation}

To evaluate the identified vulnerabilities, we evaluated our methods using real-life genomic datasets. Here, we first describe the datasets we used and then present the evaluation results.

\subsection{Datasets and Evaluation Metrics}\label{section:dataset}

We used two different genome datasets for evaluation: (i) genome dataset of CEU population from the HapMap dataset~\cite{gibbs2003international} and (ii)  OpenSNP genome dataset~\cite{opensnp}.
Using the HapMap dataset, we created the beacons and victims from CEU population which contains $164$ donors and around $4$ million SNPs for each donor. We created the correlation model (i.e., SNP-SNP relation network or similarity model) for this beacon using individuals from the same HapMap dataset that are not in the constructed beacon and set of victims.
Using the OpenSNP dataset, we created the beacons and victims from a random population which contains 2980 donors and around $2$ million SNPs for each donor. We created the correlation model using the rest of the OpenSNP dataset.

For the OpenSNP dataset, we also collected the reported phenotypes of individuals. Since sample sizes are small, we used the reported phenotypes in a binary form. From OpenSNP, we used the following commonly reported phenotypes: (i) eye color, 967 samples, (ii) hair type, 371 samples, (iii) hair color, 468 samples, (iv) tan ability, 287 samples, (v) asthma, 226 samples, (vi) lactose intolerance, 347 samples, (vii) earwax, 244 samples, (viii) tongue rolling, 434 samples, (ix) intolerance to soy, 136 samples, (x) freckling, 277 samples, (xi) ring finger being longer than index finger, 268 samples, (xii) widow peak, 176 samples, (xiii) ADHD, 154 samples, (xiv) acrophobia, 155 samples, (xv) finger hair, 155 samples, (xvi) myopia, 152 samples, (xvii) irritable bowel syndrome, 142 samples, (xviii) index finger being longer than big thumb, 131 samples, (xix) photoptarmis, 133 samples, (xx) migraine, 129 samples, and (xxi) Rh protein, 311 samples. We used $1320$ genomes which are associated with at least one of the above-mentioned phenotypes while training the models. Newly added donors are chosen from the individuals who have reported at least 10 out of 21 considered phenotypes.

We evaluated the precision and recall for the reconstruction of a victim's SNPs based on the changes in beacon responses. For precision and recall, we defined the success as correctly inferring the SNPs of the victim with at least one minor allele.
Thus, for the calculation of precision and recall, we defined (i) \textit{true positive} as correctly inferring a SNP $j$ of victim $v$ with $\hat{S}_j^v=1$ (with at least one minor allele); (ii) \textit{false positive} as incorrectly assigning $\hat{S}_j^v=1$ for $v$ who is homozygous major at that locus; (iii) \textit{true negative} as correctly inferring a SNP $j$ of victim $v$ with $\hat{S}_j^v=0$ (with no minor allele, homozygous major); and (iv) \textit{false negative} as incorrectly assigning $\hat{S}_j^v=0$ for $v$ who has at least one minor allele at that locus (i.e., heterozygous or homozygous minor).
Furthermore, we quantified the success of identifying the victim's genome among the reconstructed genomes in terms of the accuracy of the developed genotype-phenotype inference mechanism. Finally, we used a power analysis for the membership inference to show how the outcome of the genome reconstruction attack can be used for membership inference attack.

\subsection{Evaluation of Genome Reconstruction}\label{sec:eval_construction}

First, using both OpenSNP and HapMap beacons and only focusing on genome reconstruction, we evaluated and compared the baseline method (in Section~\ref{sec:baseline}) and the proposed clustering-based approach (in Section~\ref{sec:clustering}) when the size of the beacon ($n$) is $50$ and $m=m'$.
Here, we assume that the attacker can identify the victim's reconstructed genome among the other candidates. Later, we will also show that attacker can indeed identify this genome with high accuracy using public (i.e., not sensitive) phenotype information about the victim.

\begin{figure*}[t]
     \centering
     \begin{subfigure}[t]{0.49\textwidth}
         \centering
         \includegraphics[width=0.80\textwidth]{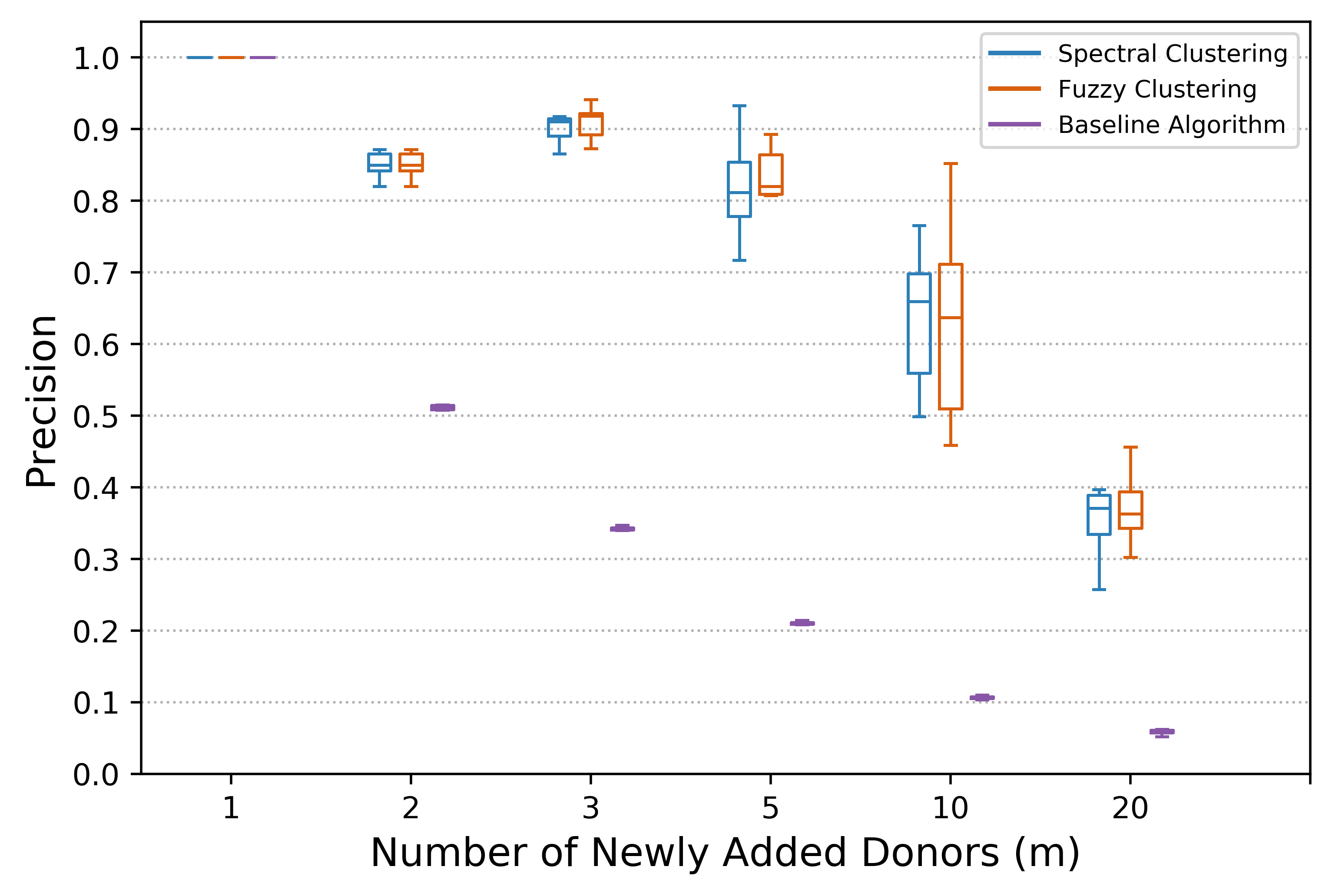}
         \caption{Precision.}
         \label{fig:NoYes-Precision}
     \end{subfigure}
     \hfill
     \begin{subfigure}[t]{0.49\textwidth}
         \centering
         \includegraphics[width=0.80\textwidth]{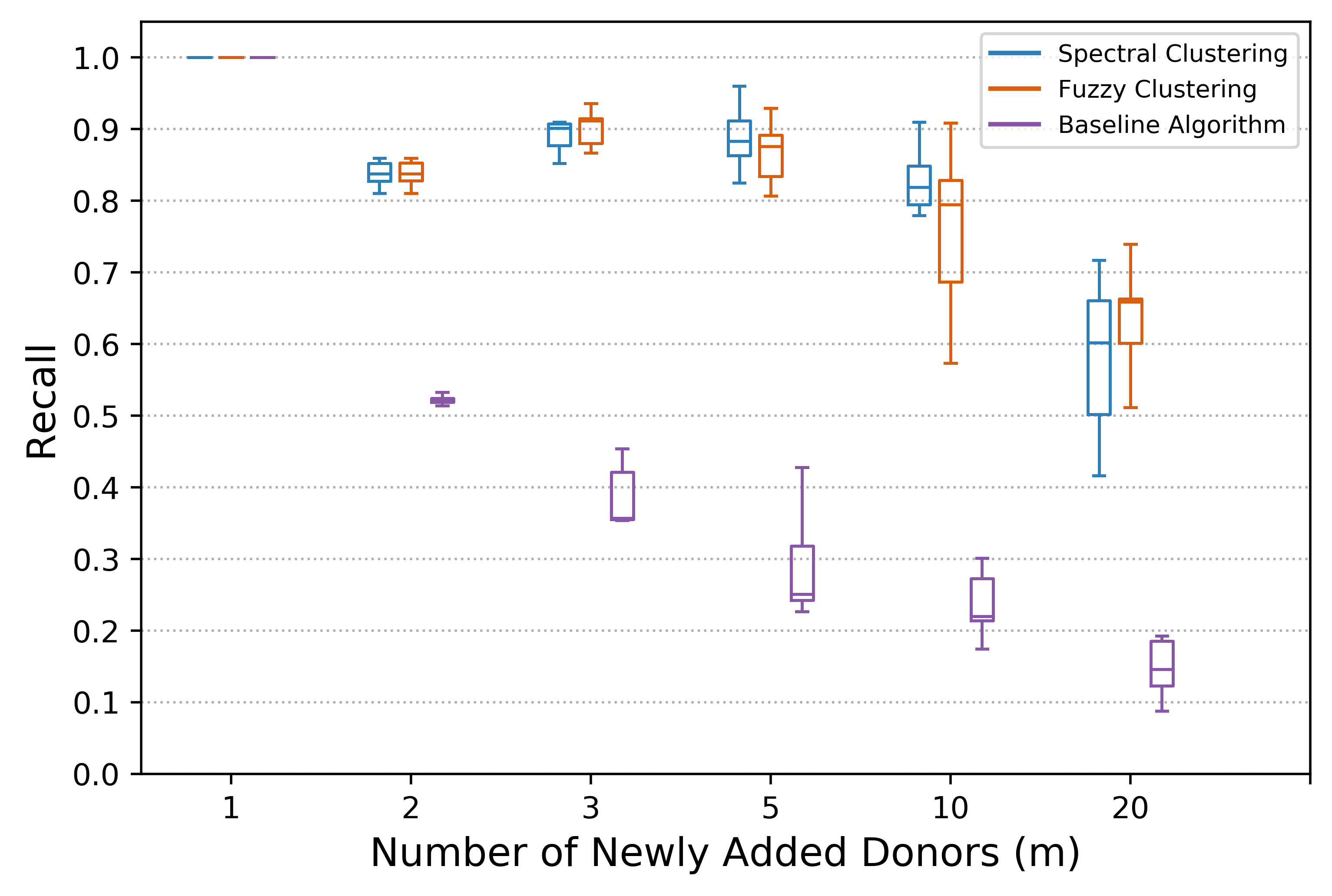}
         \caption{Recall.}
         \label{fig:NoYes-Recall}
     \end{subfigure}
     \hfill
        \caption{Precision and recall for the genome reconstruction of a newly added donor to OpenSNP beacon with varying number of newly added donors.}
        \label{fig:num_donors_opensnp}
\end{figure*}

\begin{figure*}[t]
     \centering
     \begin{subfigure}[t]{0.49\textwidth}
         \centering
         \includegraphics[width=0.80\textwidth]{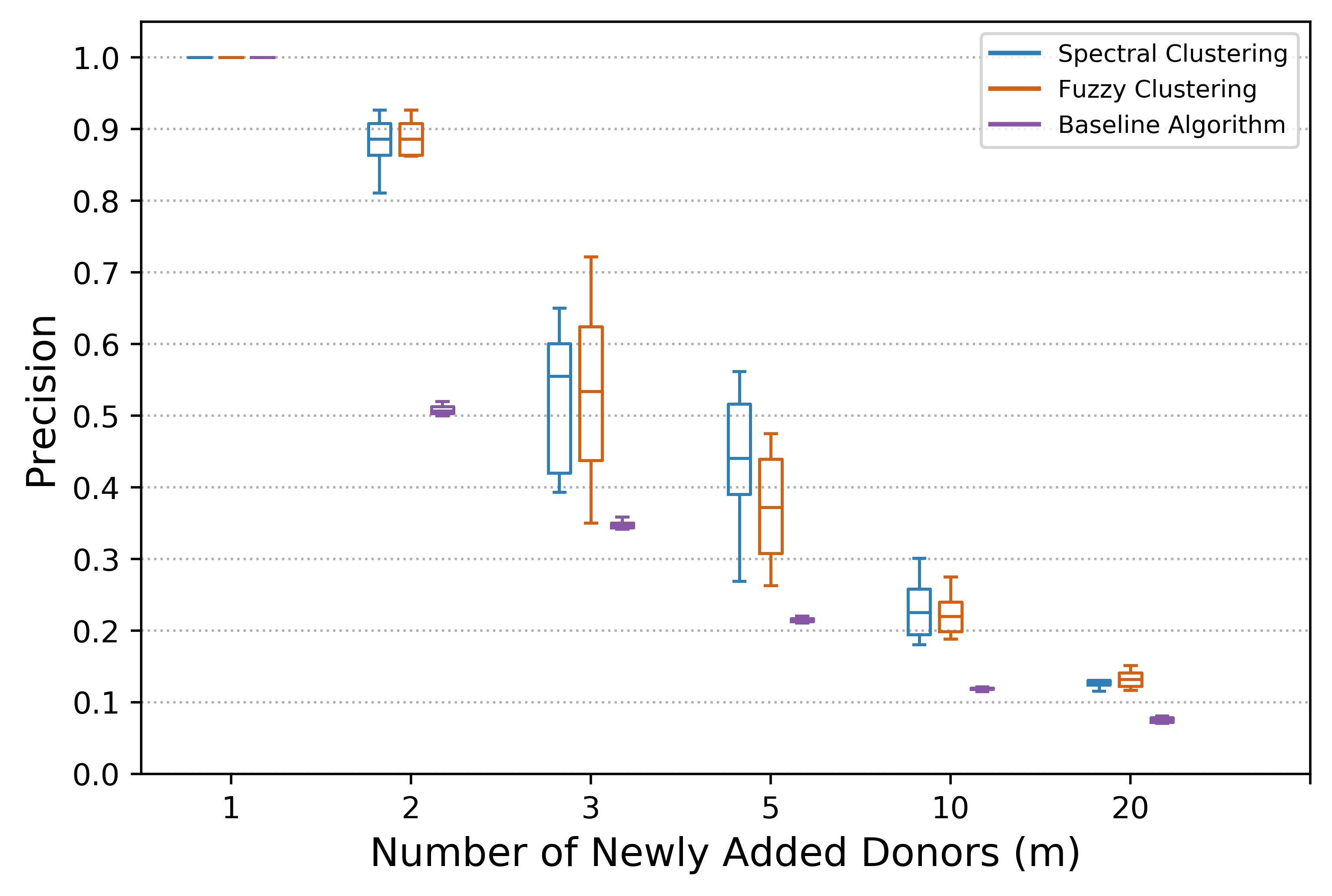}
         \caption{Precision.}
     \end{subfigure}
     \hfill
     \begin{subfigure}[t]{0.49\textwidth}
         \centering
         \includegraphics[width=0.80\textwidth]{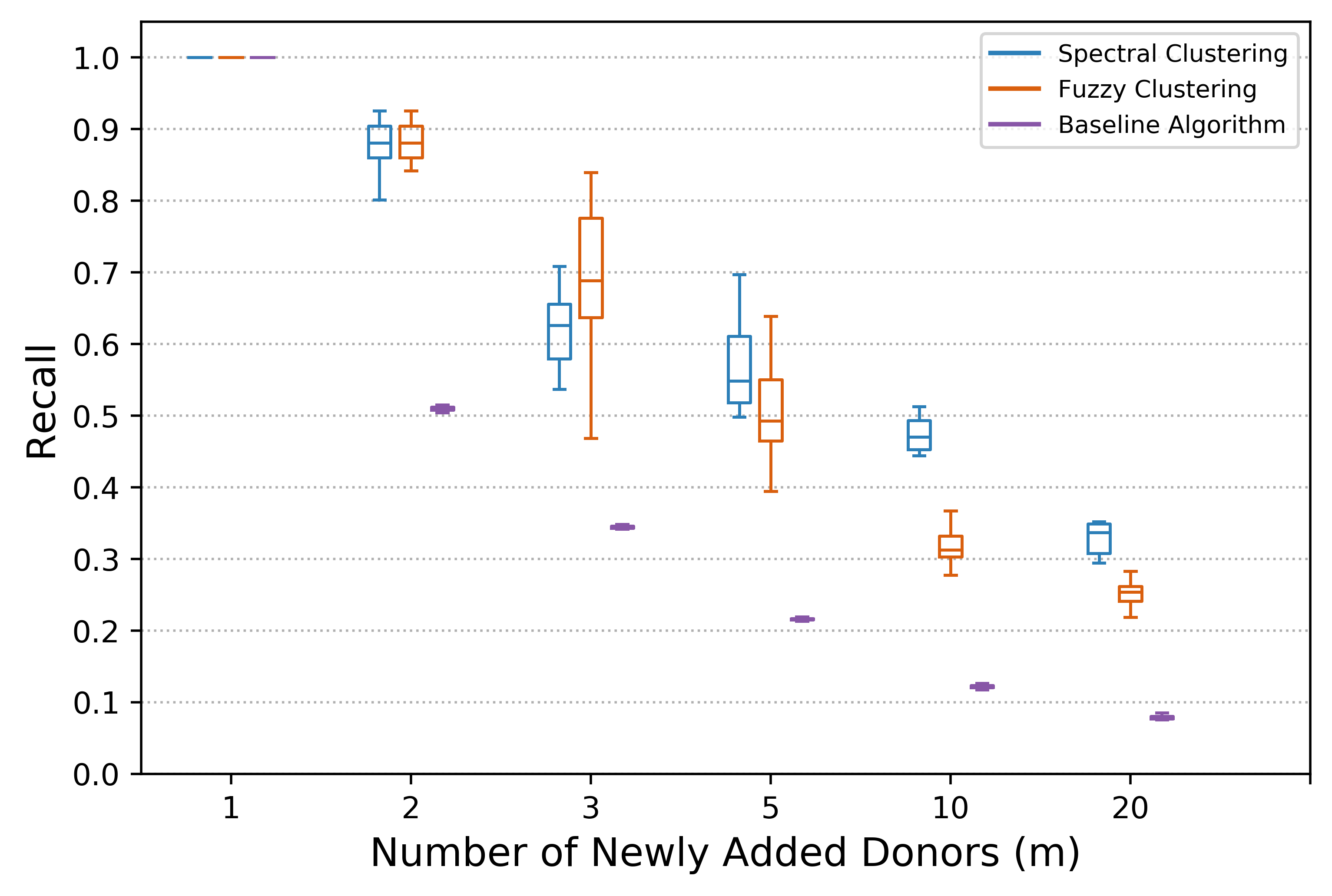}
         \caption{Recall.}
     \end{subfigure}
     \hfill
        \caption{Precision and recall for the genome reconstruction of a newly added donor to HapMap beacon with varying number of newly added donors.}
        \label{fig:num_donors_ceu}
\end{figure*}

Overall, results we obtained from both beacons are similar to each other, showing that the identified vulnerability is not dataset specific.
Figures~\ref{fig:num_donors_opensnp} and~\ref{fig:num_donors_ceu} show the precision and recall of the reconstruction for various number of newly added donors ($m$) for OpenSNP and HapMap beacons, respectively.
The results show that on average, the identified attack using spectral clustering can reconstruct the victim's genome with a precision close to $0.9$ when the size of the beacon is increased by adding $3$ people in an update. We also obtained more than $0.7$ precision and $0.8$ recall even when the size of the beacon is increased by adding $10$ people. This indicates a substantial privacy risk, especially if the reconstructed SNPs are tied to sensitive phenotypes.
Also, the baseline algorithm performs substantially worse than the proposed clustering-based approach.
The results also show that spectral clustering-based genome reconstruction is slightly better than the fuzzy clustering-based approach. We observed that allowing a SNP (that includes at least one minor allele) to be in multiple bins results in high false positives. Therefore, in the remaining of this section, we use spectral clustering-based genome reconstruction for the evaluations.

\begin{figure*}[t]
     \centering
     \begin{subfigure}[t]{0.49\textwidth}
         \centering
         \includegraphics[width=0.80\textwidth]{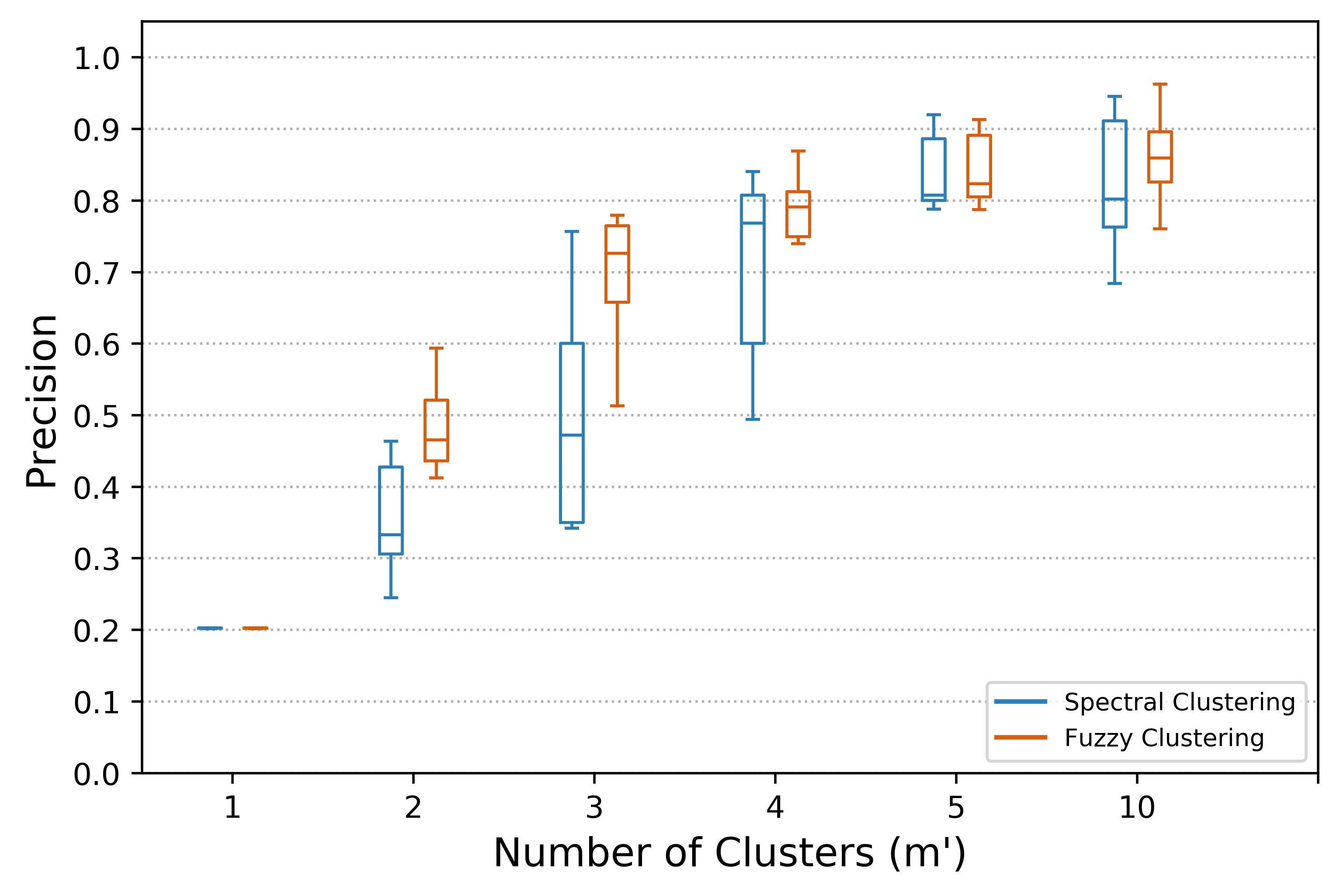}
         \caption{Precision.}
         \label{fig:vc_NoYes-Precision}
     \end{subfigure}
     \hfill
     \begin{subfigure}[t]{0.49\textwidth}
         \centering
         \includegraphics[width=0.80\textwidth]{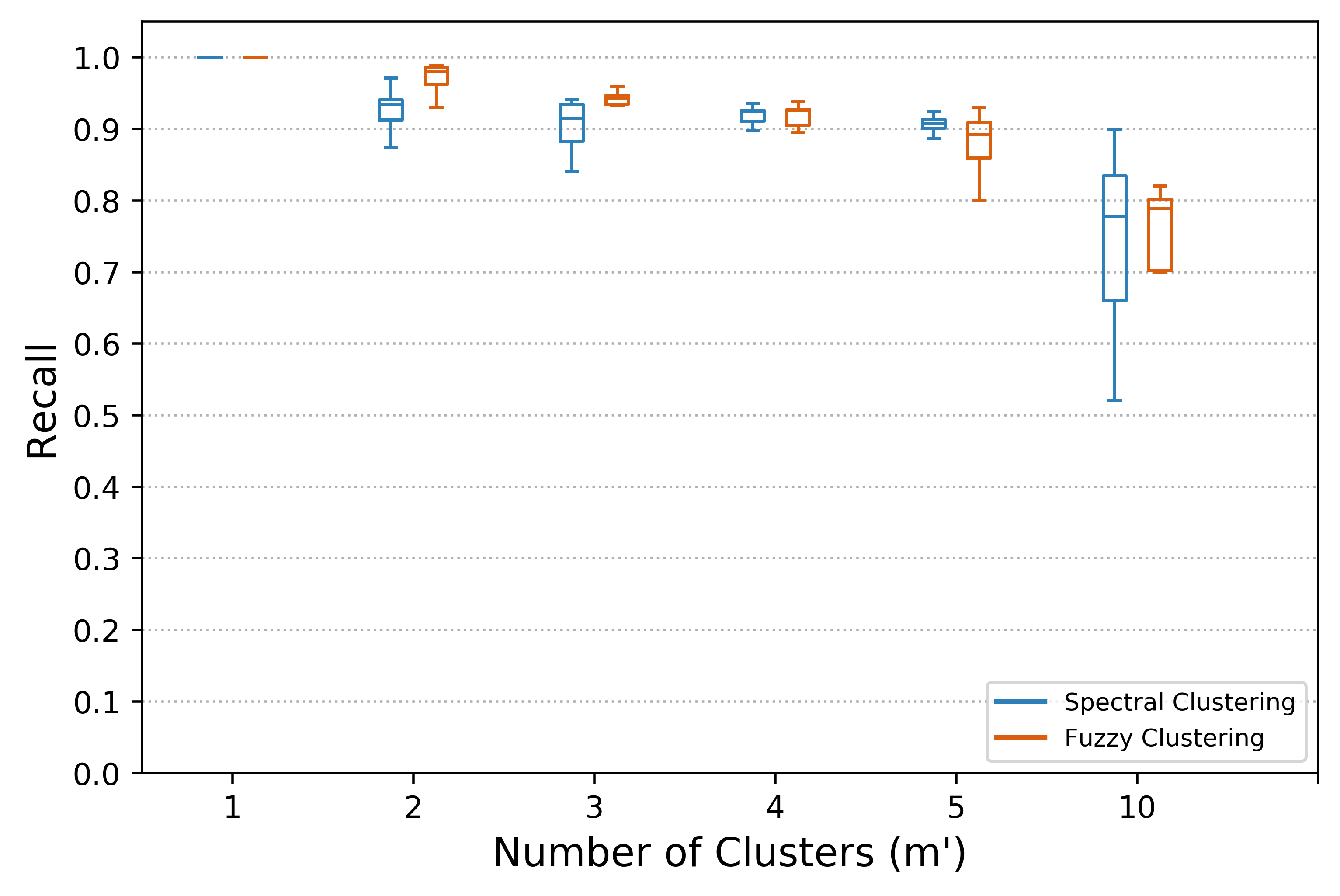}
         \caption{Recall.}
         \label{fig:vc_NoYes-Recall}
     \end{subfigure}
     \hfill
        \caption{Precision and recall for the genome reconstruction of a newly added donor to OpenSNP beacon with varying number of bins/clusters ($m'$) in the genome reconstruction attack. Number of newly added donors ($m$) is $5$.}
        \label{fig:num_clusters_opensnp}
\end{figure*}

\begin{figure*}[t]
     \centering
     \begin{subfigure}[t]{0.49\textwidth}
         \centering
         \includegraphics[width=0.80\textwidth]{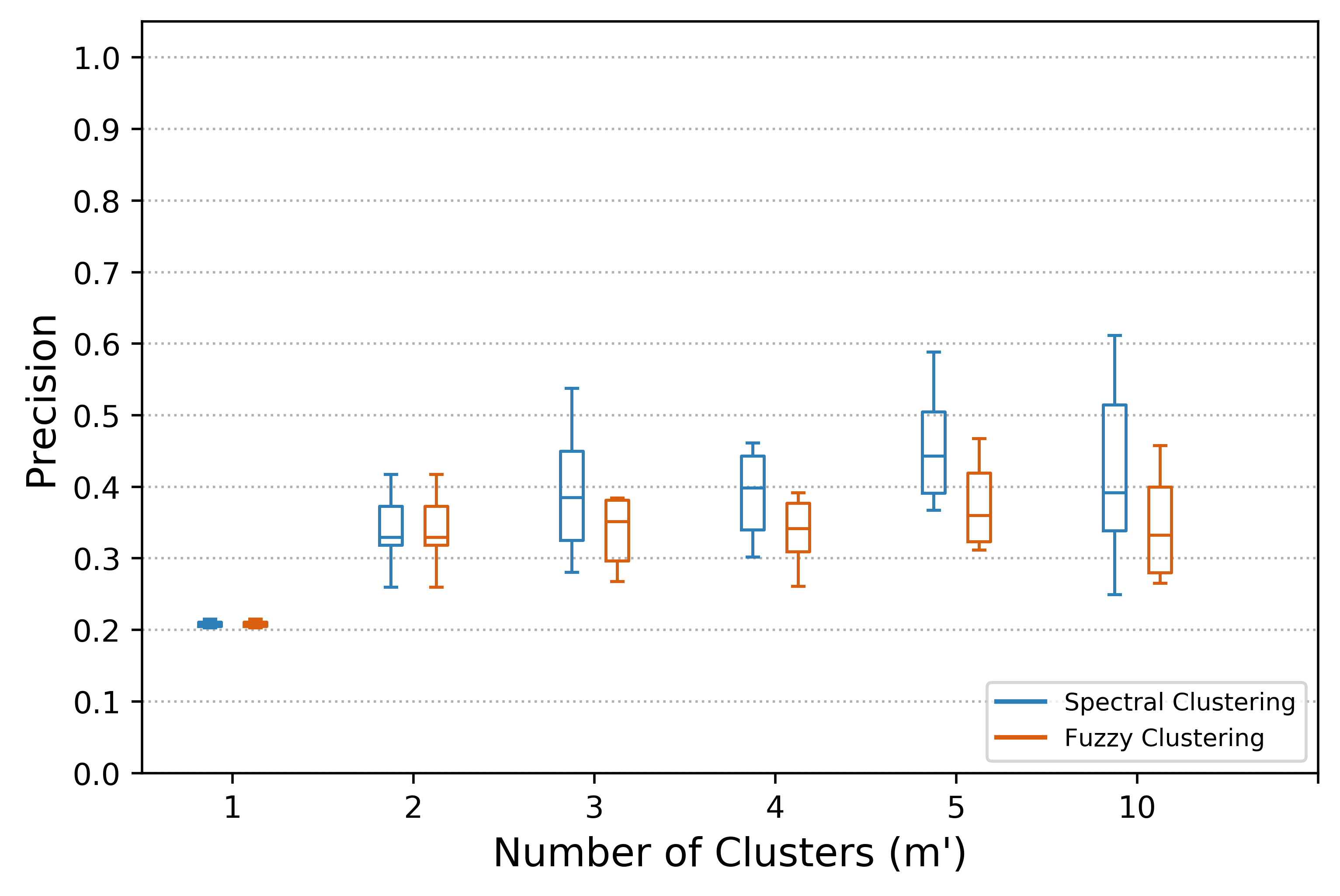}
         \caption{Precision.}
     \end{subfigure}
     \hfill
     \begin{subfigure}[t]{0.49\textwidth}
         \centering
         \includegraphics[width=0.80\textwidth]{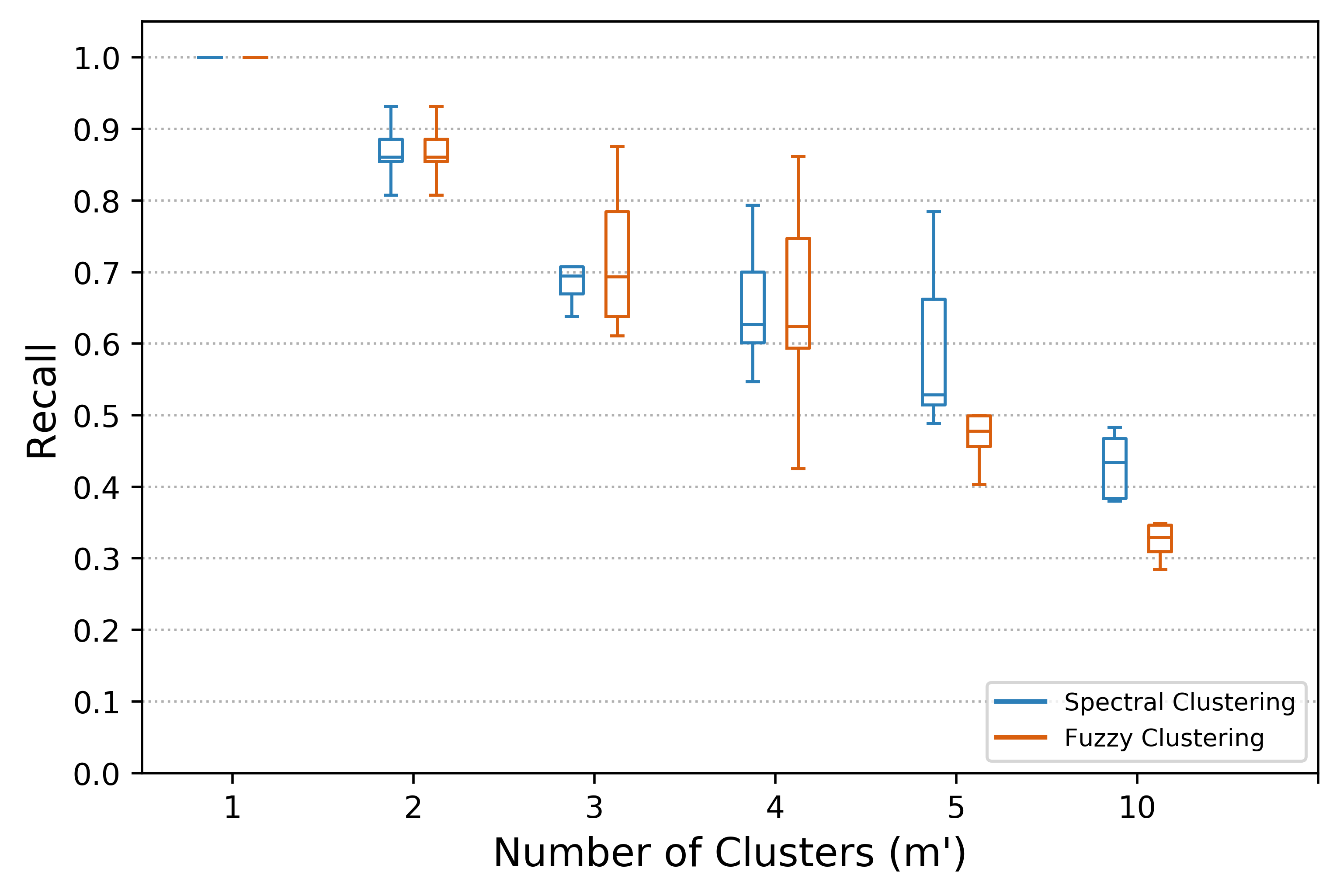}
         \caption{Recall.}
     \end{subfigure}
     \hfill
        \caption{Precision and recall for the genome reconstruction of a newly added donor to HapMap beacon with varying number of bins/clusters ($m'$) in the genome reconstruction attack. Number of newly added donors ($m$) is $5$.}
        \label{fig:num_clusters_ceu}
\end{figure*}

In Figures~\ref{fig:num_clusters_opensnp} and~\ref{fig:num_clusters_ceu}, we show the effect of varying number of bins ($m'$) in the genome reconstruction attack when the number of newly added donors $m=5$ and beacon size $n=50$ for OpenSNP and HapMap beacons, respectively. We observed that for both beacons, precision increases and recall decreases with increasing $m'$. Also, as expected, precision and recall becomes balanced when $m'=m$.

\begin{figure*}[t]
     \centering
     \begin{subfigure}[t]{0.49\textwidth}
         \centering
         \includegraphics[width=0.80\textwidth]{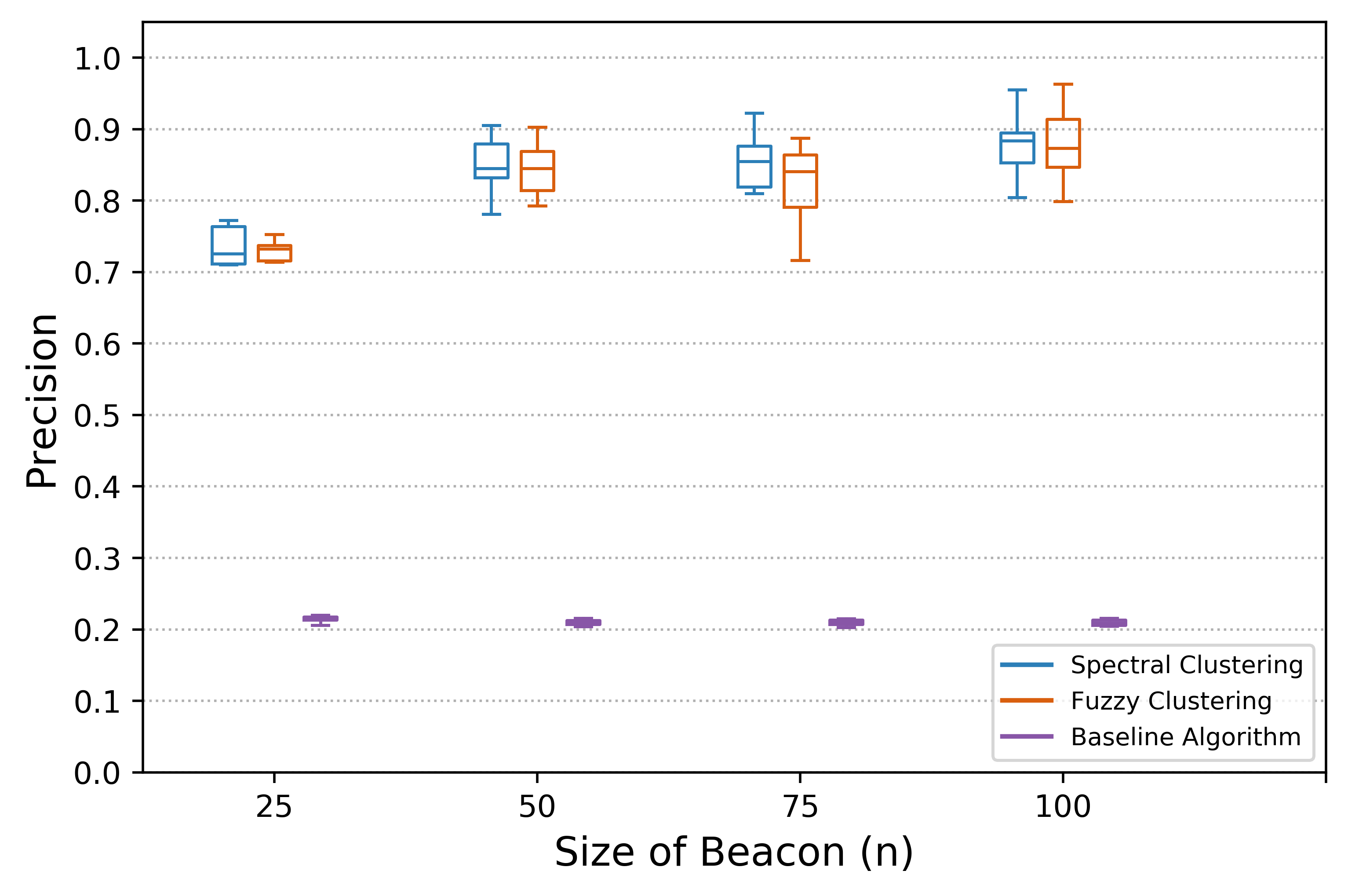}
         \caption{Precision.}
         \label{fig:beacon_size_NoYes-Precision}
     \end{subfigure}
     \hfill
     \begin{subfigure}[t]{0.49\textwidth}
         \centering
         \includegraphics[width=0.80\textwidth]{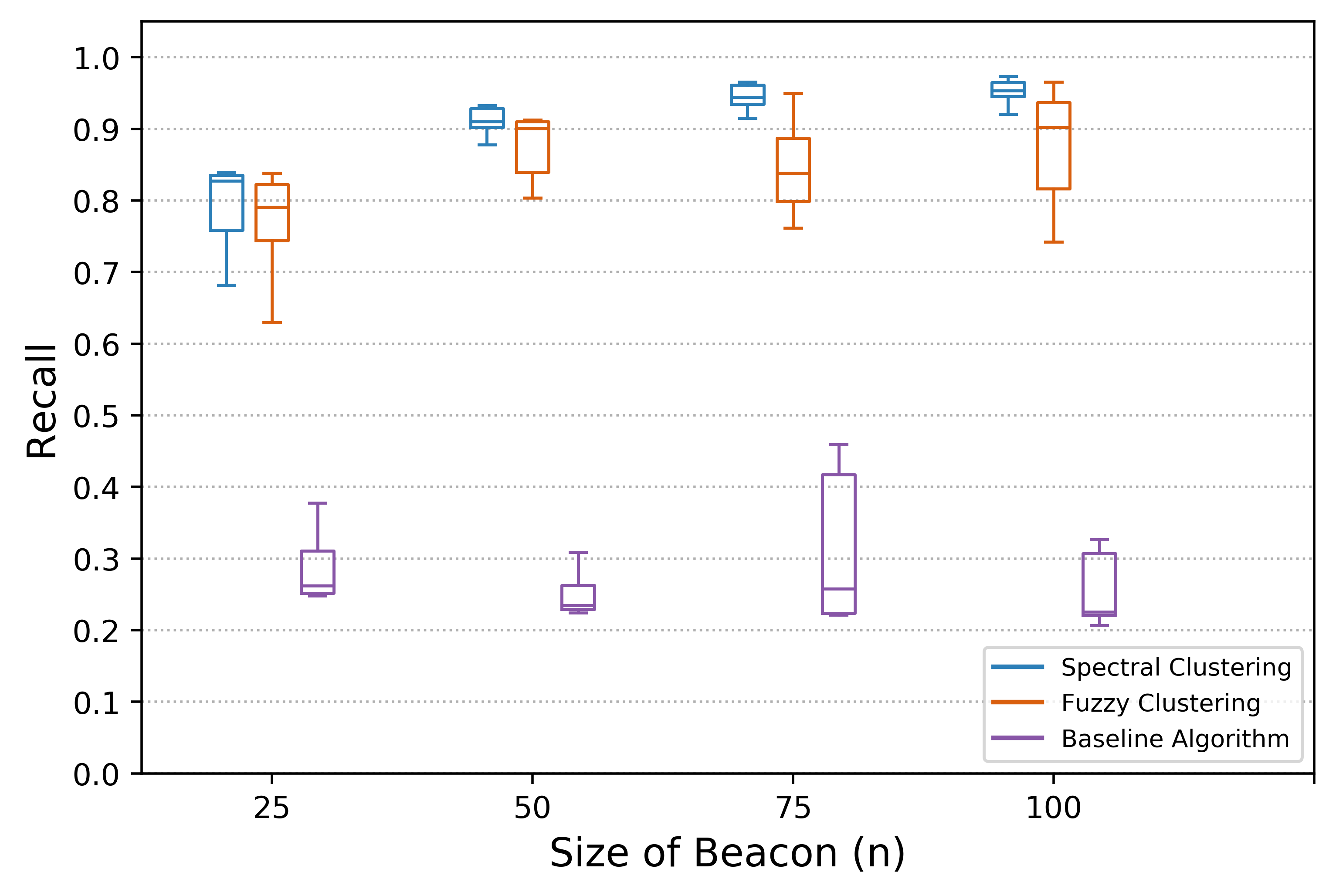}
         \caption{Recall.}
         \label{fig:beacon_size_NoYes-Recall}
     \end{subfigure}
     \hfill
        \caption{Precision and recall for the genome reconstruction of a newly added donor to OpenSNP beacon with varying number of beacon size ($n$). Number of newly added donors $m$ is $5$ and $m'=m$ for all plots.}
        \label{fig:beacon_size_opensnp}
\end{figure*}

\begin{figure*}[t]
     \centering
     \begin{subfigure}[t]{0.49\textwidth}
         \centering
         \includegraphics[width=0.80\textwidth]{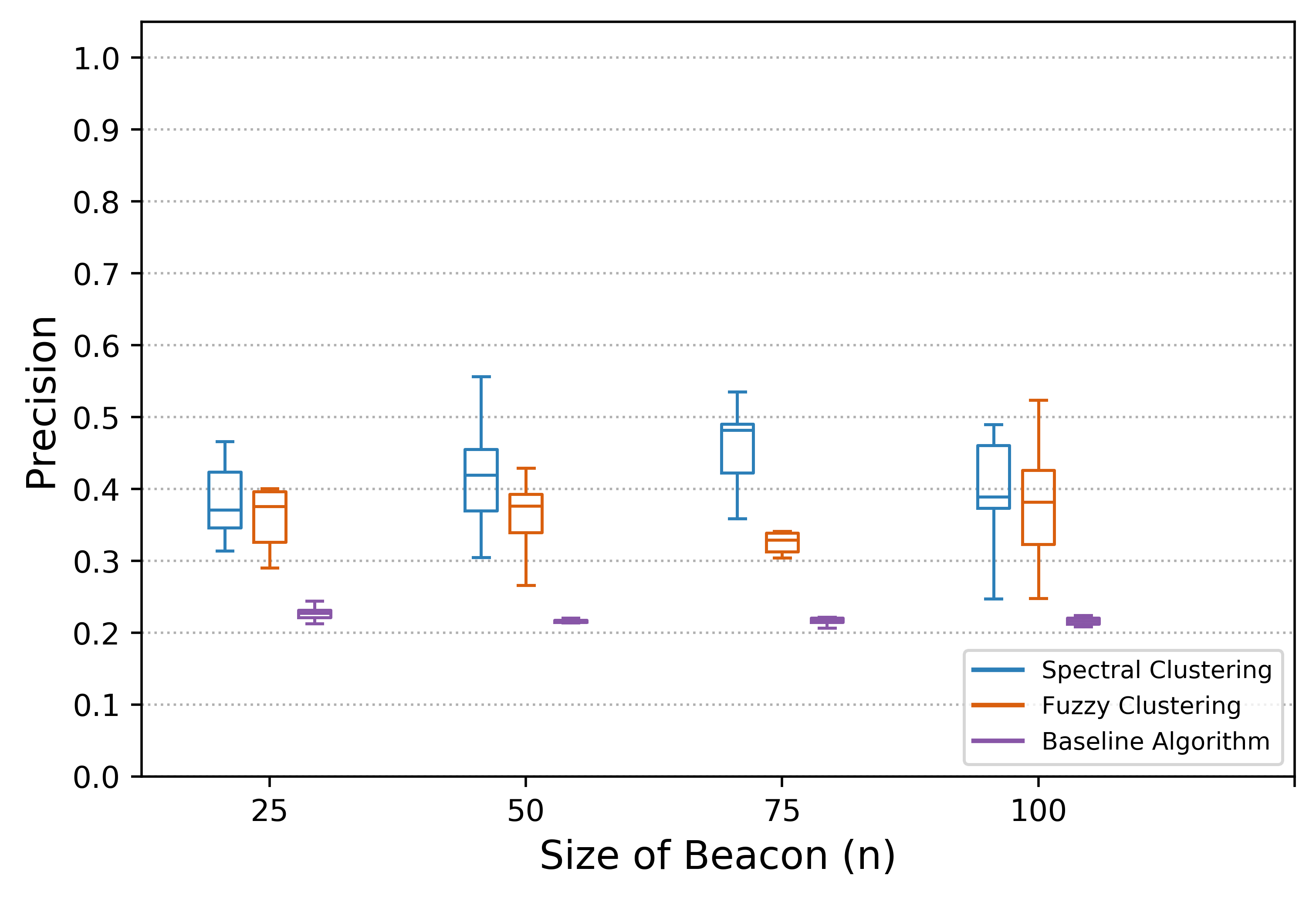}
         \caption{Precision.}
     \end{subfigure}
     \hfill
     \begin{subfigure}[t]{0.49\textwidth}
         \centering
         \includegraphics[width=0.80\textwidth]{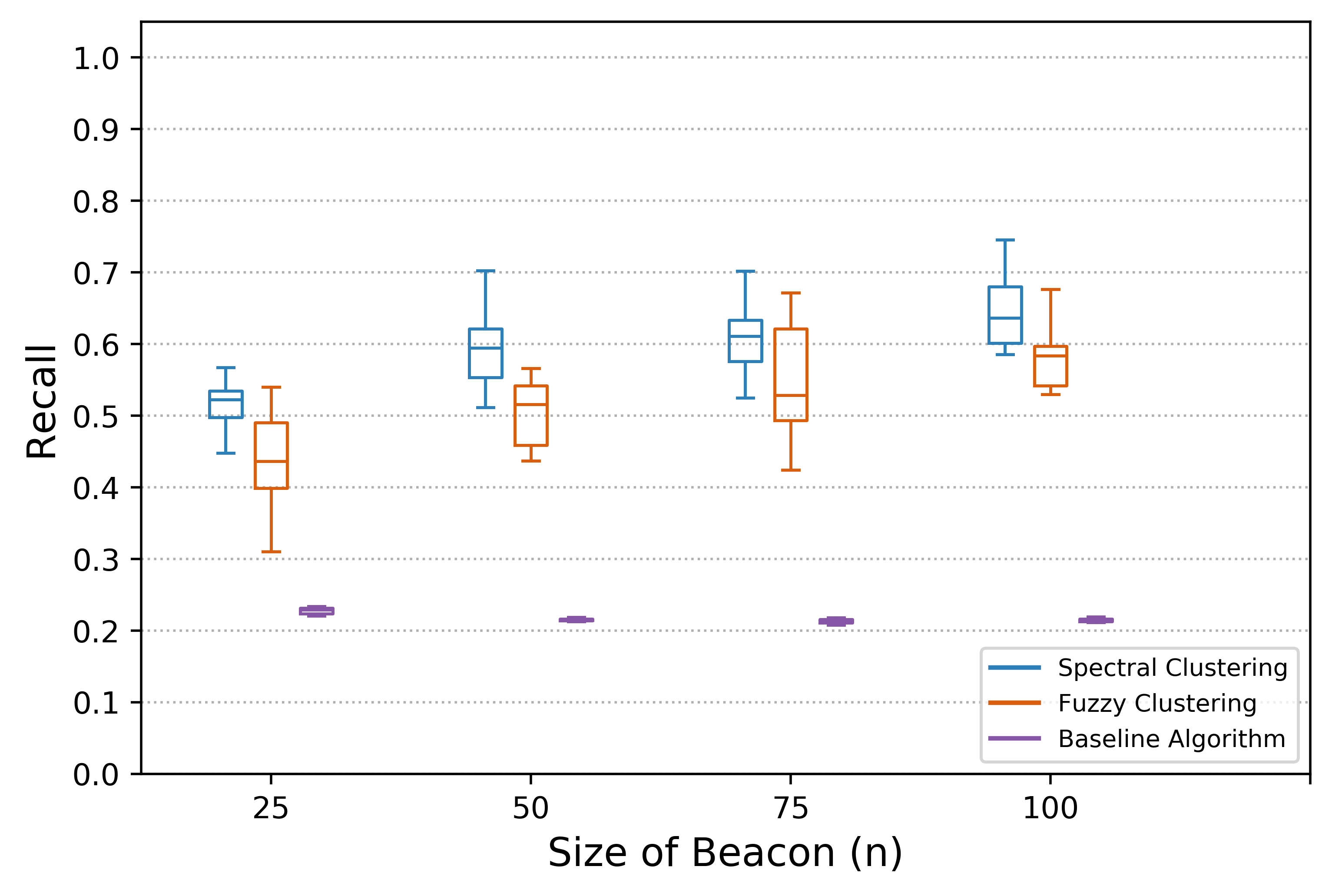}
         \caption{Recall.}
     \end{subfigure}
     \hfill
        \caption{Precision and recall for the genome reconstruction of a newly added donor to HapMap beacon with varying number of beacon size ($n$). Number of newly added donors $m$ is $5$ and $m'=m$ for all plots.}
        \label{fig:beacon_size_ceu}
\end{figure*}

Next, in Figures~\ref{fig:beacon_size_opensnp} and~\ref{fig:beacon_size_ceu}, we show the effect of the beacon size ($n$) at time $t$ when $5$ new donors are added between times $t$ and $t+\delta$ for OpenSNP and HapMap beacons, respectively. Here, we assume that the number of bins ($m'$) is equal to the number of newly added donors ($m$).
We observed that as the size of the beacon increases, both the precision and recall of the reconstruction attack slightly increases (for a fixed number of newly added donors). This shows that the success of the identified attack mainly relies on the fraction of the newly added donors to the beacon, and it is independent of the size of the beacon at time $t$. Note that even the success of the genome reconstruction is high, the number of flipped responses (from ``no'' to ``yes'') decreases when beacon size is increased. This might result in lower performance in phenotype inference and membership inference parts of the attack.



\subsection{Identifying the Victim's Genome Using Phenotype Inference}\label{sec:eval_phenotype}

Here, we evaluate the success of the attacker in identifying the reconstructed genome of the victim among all reconstructed genomes using the algorithm in Section~\ref{sec:phenotype}. Since HapMap dataset does not include phenotype information about the genome donors, we only use the OpenSNP beacon for this evaluation.

We employed and compared several machine learning models for genotype-phenotype associations, including: Logistic Regression~\cite{logistic}, SVM~\cite{svm}, Multi-layer Perceptron~\cite{mlp}, Random Forest~\cite{randomforest}, and XGBoost~\cite{xgboost}. Among these, we obtained the highest classifier accuracy with the Random Forest, and hence all reported results are based on this model.

\begin{figure}[h]
\centering
\includegraphics[width=0.45\textwidth]{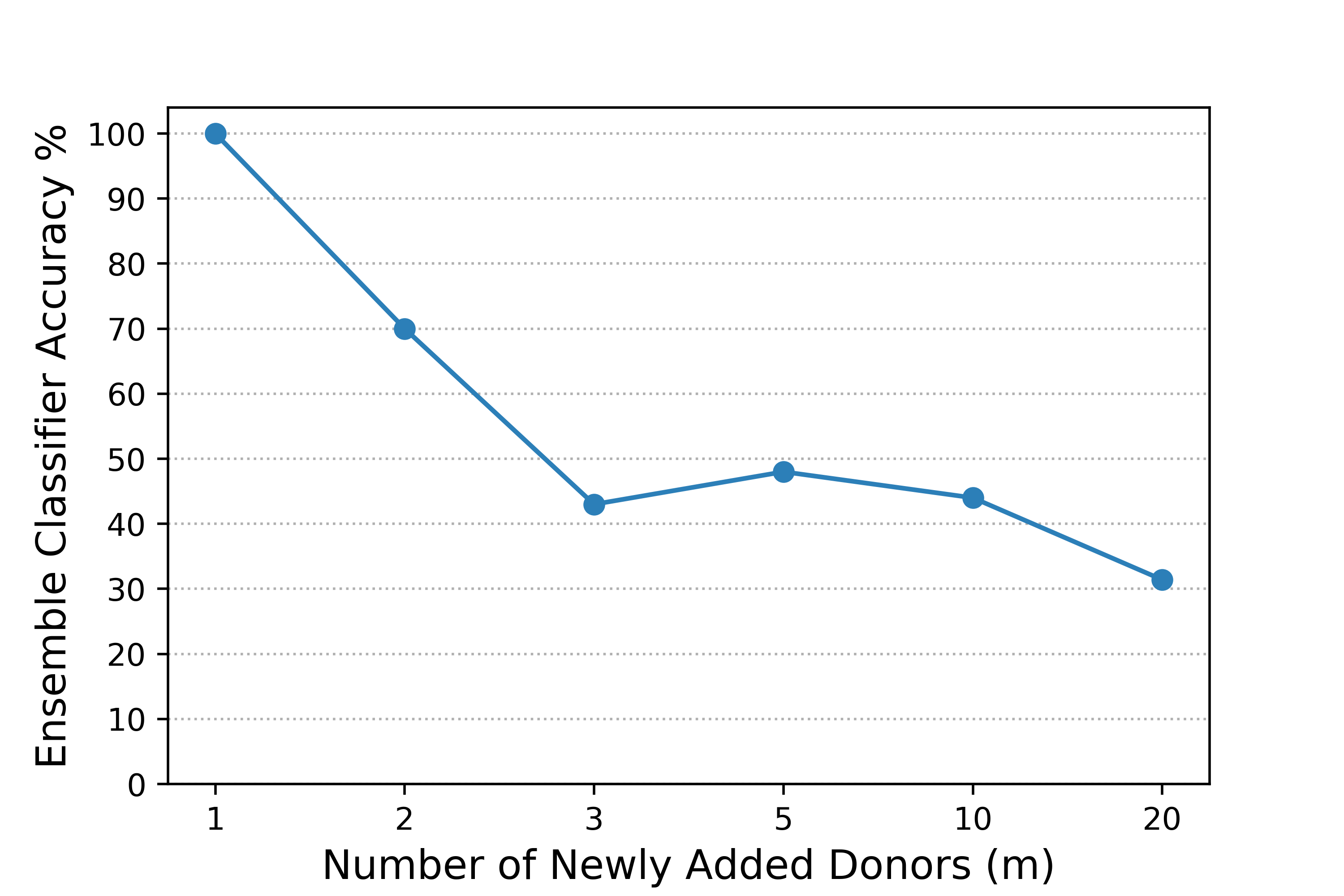}
\caption{Classification accuracy of genotype inference from phenotype for varying number of newly added donors ($m$) to the beacon.}
\label{fig:phenotype_accuracy}
\end{figure}

In Figure~\ref{fig:phenotype_accuracy}, we show the ensemble classifier accuracy for varying number of newly added donors to the beacon (here, we assumed $m'=m$ and we observed similar patterns when $m'\neq m$ as well). We used the original genomes of individuals in the training dataset when building the model. For test, we used reconstructed genomes of the victims (that may have noise due to reconstruction error). Beacon size is 50 in these experiments (i.e., $n=50$).

We observed that the proposed algorithm provides $70\%$ accuracy when the size of the beacon is increased by adding $2$ individuals in the update, and the accuracy slightly decreases with increasing number of newly added donors. These results show that the attacker can identify the reconstructed genome of the victim among all $m'$ reconstructed genomes with high accuracy.

\subsection{Using Genome Reconstruction in Membership Inference}\label{sec:eval_membership}

In Section~\ref{sec:eval_construction}, we evaluated the success of the reconstruction and in Section~\ref{sec:eval_phenotype}, we showed that the attacker is able to identify the victim among many added donors with high accuracy. Here, we show a severe consequence of the proposed genome reconstruction attack, in which the outcome of the previous steps can be utilized in a membership inference attack. For this, we randomly constructed two non-overlapping beacons from the OpenSNP dataset: (i) $B_1$ includes 50 individuals and (ii) $B_2$ includes 60 individuals. We assume that $B_2$ is associated with a privacy-sensitive phenotype and the goal of the attacker is to infer the membership of the victim to $B_2$.
We also assume that $m$ new individuals are added to $B_1$ at time $t+\delta$ and the victim is among these newly joined donors. The attacker only knows that the victim is among these $m$ individuals that are added to $B_1$ at time $t+\delta$ along with a snapshot of $B_1$ at time $t$.

First, we applied the spectral clustering-based genome reconstruction (that provides the best performance in Section~\ref{sec:eval_construction}) to reconstruct the genomes of newly joined $m$ donors to $B_1$. Then, we identified the reconstructed genome of the victim using phenotype information about the victim (as in Section~\ref{sec:phenotype}). Finally, using the reconstructed genome of the victim, we conducted the membership inference attack on $B_2$ using the Optimal attack (as described in Section~\ref{sec:membership_inference}).

We used the identification accuracy in Section~\ref{sec:phenotype} to construct and infer victims' genomes for alternate and null hypotheses. For instance, when $m=2$ we have $70\%$ identification accuracy. In this scenario, $14$ genomes are chosen from correctly reconstructed genomes, while the remaining $6$ genomes are chosen from incorrectly reconstructed genomes for corresponding victims.

\begin{figure}[h]
\centering
\includegraphics[width=0.45\textwidth]{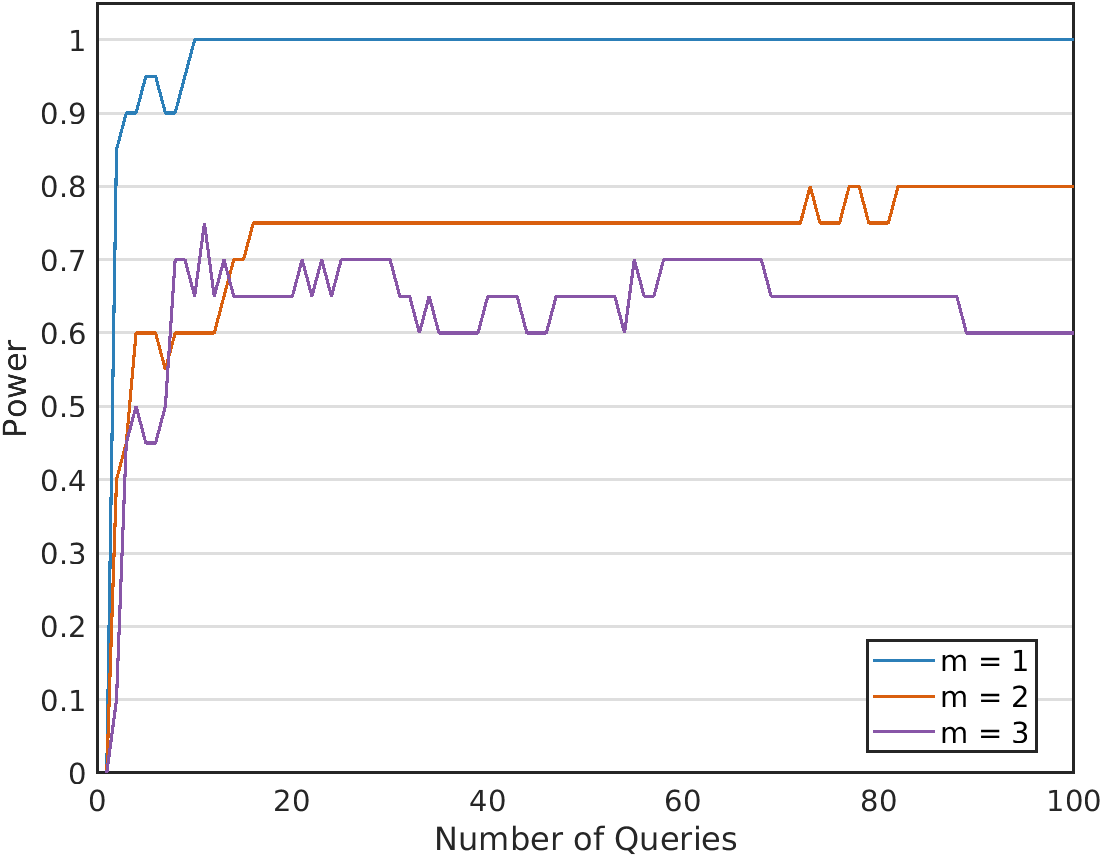}
\caption{Power of membership inference attack on beacon $B_2$ with varying number of newly added donors ($m$) to beacon $B_1$.}
\label{fig:inference_m}
\end{figure}

In Figure~\ref{fig:inference_m}, we show the power plots of this attack with varying number of newly added donors ($m$) to beacon $B_1$. As expected, with decreasing values of $m$, the power increases faster since the accuracy of genome reconstruction increases (and hence the error rate of the membership inference attack decreases). For instance, when the victim is the only newly added donor to beacon $B_1$ ($m=1$), the attacker can reconstruct their genome and then infer the victim's membership to beacon $B_2$ with a very high confidence (100\% power) in just slightly more than $15$ queries. We also observed that when $m$ is increased, the power decreases, yet still reaches to $0.8$ with approximately $80$ queries when $2$ individuals are added. These results show that the attacker may confidently conduct membership inference attacks as a result of genome reconstruction even though it has many sources of uncertainties in its input for membership inference.

%% file: sections/discussion.tex
\section{Discussion}\label{section:discussion}

Beacons have been widely accepted by the community as the best standard for ease of set up and encouraging collaboration without compromising security. However, the privacy pitfalls have cast doubts on their usability because of possible membership inference attacks.
This work pinpoints a new information leak and identifies beacon updates as a new risk, which leads to genome reconstruction attacks. We show that an attacker can efficiently and accurately link this new vulnerability to a membership inference attack. Our results show that a beacon admin should never add a single donor to the beacon. Even when two or more people are added, the attacker still has power to launch a subsequent membership inference attack. We also show that the reconstruction has high and balanced precision and recall in this case. We recommend that beacon updates are done in large batches, which decreases the utility of reconstruction and downstream attacks.

In the following subsections, we discuss some alternative scenarios for the proposed attack, practical use case of the identified vulnerability, and potential mitigation techniques.

\subsection{Donors Leaving the Beacon}\label{sec:leaving_donors}

In Sections~\ref{section:proposed_work} and~\ref{section:evaluation}, we presented and evaluated the identified vulnerability by only considering the newly joined donors to the beacons. It is also possible that existing donors may leave the beacon. However, such a scenario can be easily addressed by using the identified attack mechanism. Considering the donors that leave the beacon brings up two different scenarios: (i) victim is among the newly joined donors (while there are also donors leaving the beacon between times $t$ and $t+\delta$) and (ii) victim is among the donors that leave the beacon (while there may be other donors leaving or joining the beacon between times $t$ and $t+\delta$).

Scenario in (i) is no different than what we discussed in Section~\ref{section:proposed_work}. The number of ``no'' responses at time $t$ that turn to ``yes'' at time $t+\delta$ does not change due to the donors leaving the beacon. On the other hand, some ``yes'' responses at time $t$ may turn to ``no'' at time $t+\delta$ due to the donors leaving the beacon. However, such responses do not provide information about the minor alleles of the victim, and hence we do not consider such responses in this work. In scenario (ii), ``yes'' responses at time $t$ that turn to ``no'' at time $t+\delta$ will provide information about the minor alleles of the victim (and other donors that leave the beacon during that time interval). Using such responses, one will need to run the algorithms proposed in Section~\ref{section:proposed_work} to reconstruct the genome of the victim. Thus, it is trivial to consider both newly joining and leaving donors in the proposed attack mechanism.

\subsection{Risk Quantification for the Genome Reconstruction Attack}\label{sec:usecase}

The identified vulnerability and the proposed attack algorithm can be used as a privacy risk quantification tool by the beacon operator. For this, we foresee a simulation-based technique to quantify the risk and show it to the beacon operator. This will be a customized technique for each donor in the beacon and the following discussion is for one particular donor. Assume that a total of $m$ new donors are gathered by the beacon between times $t$ and $t+\delta$. To quantify the genome reconstruction risk, one may run the attack we introduced in Section~\ref{section:proposed_work}, pretending the donor is added to the beacon along with the other $(m-1)$ newcomer donors and compute the fraction of the SNPs that can be reconstructed. Then, using public sources (such as HapMap), one can gather a small number (e.g., $s$) of genomes belonging to individuals from the same population as the donor. Then, the same attack can be run for the selected $s$ people (i.e., adding each random individual along with the other $(m-1)$ newcomer donors), their reconstruction rates can be set as the baseline, and eventually, a privacy risk percentile can be provided for the donor. Moreover, for all correctly inferred SNPs, one can perform a pathogenic scan on ClinVar~\cite{landrum2017clinvar} to inform the donor about what traits they might be linked should their genome is put onto the beacon. Using this information and based on the privacy risk of the donor, either the donor or the beacon operator will decide whether or not to add the donor to the beacon at time $t+\delta$. This process can be repeated for all the newcomer donors.

We foresee that using such a quantification algorithm, a potential beacon participant can provide informed consent about how (and what portion of) their data can be used by the beacons (e.g., when the beacon can start using their data in its responses or when the beacon should stop using their data). Similarly, such a tool can guide a beacon operator on the number of participants to include in a batch to update the beacon.

\subsection{Mitigation Techniques}\label{sec:mitigation}

To mitigate membership inference attacks against beacons, several countermeasures have been proposed. Shringarpure and Bustamante considered: (i) increasing the beacon size, (ii) sharing only small genomic regions, (iii) using single population beacons, (iv) not publishing the metadata of a beacon, and (v) adding control samples to the beacon dataset~\cite{shringarpure2015privacy}. Raisaro \emph{et al.} proposed assigning a query budget for each individual's genome as a countermeasure. However, later von Thenen \emph{et al.} showed that such query budgets are not effective considering the auxiliary information of the attacker about the victim and correlations between the SNPs~\cite{vonthenen2018reidentification}. Lately, Al Aziz \emph{et al.} proposed two algorithms that are based on randomizing the response set of the beacons with the goal of protecting beacon members' privacy while maintaining the efficacy of the beacon servers~\cite{al2017aftermath}. However, most of such techniques directly reduce the utility of the beacon without carefully analyzing a balance between privacy (of beacon participants) and utility (of beacon responses). Thus, we believe that existing countermeasures proposed for membership inference are not directly applicable to mitigate genome reconstruction attack.

To mitigate genome reconstruction, here we suggest three simple methods: (i) updating the beacon content when $m$ > 1; (ii) adding (or removing) donors after quantifying their risks against genome reconstruction (as discussed in Section~\ref{sec:usecase}); and (iii) adjusting diversity of the beacon to have beacons with mixed ethnicity genome donors. We observed that for beacons with mixed ethnicity donors, it is hard to construct the correlation model (unless the beacon discloses the ethnicities of the donors as metadata), and hence it is hard to conduct the proposed correlation-based genome reconstruction attacks. We will further work on more sophisticated countermeasures in future work.

%% file: sections/conclusion.tex
\section{Conclusion and Future Work}\label{section:conclusion}

Thus far, the only privacy vulnerability that has been identified for beacons was membership inference. We have identified and, via extensive analysis, showed the impact of another serious privacy concern for beacons: genome reconstruction. We showed the practicality of the identified privacy concern in real-life by showing the whole attack strategy including genotype-phenotype inference. Furthermore, we showed how genome reconstruction attack can be used together with the membership inference to identify privacy-sensitive phenotypes of individuals.
In future work, we will develop privacy-risk quantification tools for beacon operators (and donors) using the identified vulnerability and also considering the risk of membership inference.
Furthermore, we will work on mitigation techniques for the identified vulnerability while preserving the utility of beacon content and beacon responses. 